\documentclass[fleqn,usenatbib]{mnras}

\usepackage{newtxtext,newtxmath}

\usepackage[T1]{fontenc}

\DeclareRobustCommand{\VAN}[3]{#2}
\let\VANthebibliography\thebibliography
\def\thebibliography{\DeclareRobustCommand{\VAN}[3]{##3}\VANthebibliography}

\usepackage{svg}
\usepackage{graphicx}	
\usepackage{amsmath}	
\usepackage{subcaption}
\usepackage{courier}

\newcommand{\kms}{\,km s$^{-1}$} 

\title[Superadiabaticity and the HD limit]{Superadiabaticity and the metallicity independence of the Humphreys-Davidson limit}

\author[G N Sabhahit et al.]
{Gautham N. Sabhahit,
Jorick S. Vink, 
Erin R. Higgins, 
Andreas A. C. Sander
\\
\\
$^{1}$Armagh Observatory and Planetarium, College Hill, Armagh, BT61 9DG, Northern Ireland
}

\date{Accepted XXX. Received YYY; in original form ZZZ}

\pubyear{2015}

\begin{document}
\label{firstpage}
\pagerange{\pageref{firstpage}--\pageref{lastpage}}
\maketitle

\begin{abstract}
\noindent
The Humphreys-Davidson (HD) limit sets the boundary between evolutionary channels of massive stars that either end their lives as red supergiants (RSGs) or as the hotter blue supergiants (BSGs) and Wolf-Rayet stars. Mixing in the envelopes of massive stars close to their Eddington limit is crucial for investigating the upper luminosity limit of the coolest supergiants. We study the effects of excess mixing in superadiabatic layers that are dominated by radiation pressure, and we critically investigate the effects of mixing and mass loss on the evolution of RSGs with $\log (T_{\text{eff}}/\mathrm{K}) < 3.68$ -- as a function of metallicity. Using MESA, we produce grids of massive star models at three metallicities: Galactic $(Z_\odot)$, LMC $(\frac{1}{2}Z_\odot)$ and SMC $(\frac{1}{5}Z_\odot)$, with both high and low amounts of overshooting to study the upper luminosity limit of RSGs. We systematically study the effects of excess mixing in the superadiabatic layers of post-main sequence massive stars, overshooting above the hydrogen core and yellow supergiant (YSG) mass-loss rates on the fraction of core helium burning time spent as a RSG. We find that the excess mixing in the superadiabatic layers is stronger at lower metallicities, as it depends on the opacities in the hydrogen bump at $\log (T_{\text{eff}}/\mathrm{K}) \approx 4$, which become more pronounced at lower metallicity.  This shifts the cutoff luminosities to lower values at lower metallicities, thus balancing the first-order effect of mass loss. The opposing effects of mass loss and excess envelope mixing during  post-main sequence evolution of stars with higher overshooting potentially results in a metallicity-independent upper luminosity limit. 
\end{abstract}




\section{Introduction}

In this paper, we investigate the stellar upper luminosity limit \citep{HD1979} at various metallicities $Z$ with detailed stellar evolution mixing experiments for superadiabaticity in radiatively-dominated stellar envelopes close to the Eddington limit. 

Ever since the fact that red supergiants (RSGs) were not detected above a luminosity threshold $\log (L/L_\odot) \approx 5.8$, it has been assumed this general absence of RSGs was the result of strong stellar wind mass loss in close proximity to the Eddington limit \citep{Lamers1988}. 
The challenge in explaining the evolution of the most massive stars is related to the physics of the Eddington limit. This could lead to boosted mass-loss rates \citep{Vink2011, Best2014}, and possibly the inflation of stellar envelopes \citep{Ishii1999, Graf2012, Gras2021}, although 3D simulations suggest inflation may be inhibited \citep{Jiang2015}.

The uncertain single-star physics of stellar envelopes is of key relevance to understanding both the HD limit as well as luminous blue variable (LBV) S Doradus variations \citep{Gras2021}. Furthermore, it has direct implications on binary evolution involving the interaction probability of close binary systems and gravitational wave (GW) events \citep[e.g.][]{Klencki2021}. Moreover, this physical phenomenon lies at the heart of the distribution of hydrogen (H) rich type II supernovae (SNe) versus H-poor type Ibc SNe. 
A better understanding of supergiant envelopes is also critical for understanding the maximum black hole (BH) mass of blue supergiant (BSG) models up to approx. 85 $M_\odot$ \citep{Vink2020}.

For decades, it was assumed that the critical luminosity of the HD limit would grow at lower $Z$ due to the smaller mass-loss rates of massive stars at low $Z$ \citep{Abbott1982, Vink2001}, and it was quite a surprise that \citet{Davies2018} re-evaluated the HD limit in the low $Z$ Small Magellanic Cloud (SMC) to a value of $\log (L/L_\odot) \approx 5.5$, similarly low as that of the twice higher $Z$ of the Large Magellanic Cloud (LMC).

In order to explain both the high $Z$ Galactic HD limit of $\log (L/L_\odot) \approx 5.8$ and the lower \citet{Davies2018} HD limit of $\log (L/L_\odot) \approx 5.5$ for the LMC and SMC, \citet{Higgins2020} showed that this was feasible with very efficient semiconvection \citep[see also][]{Abel2019} albeit for a low amount of core overshooting. \citet{Davies2018} had earlier suggested that the $Z$-independent HD limit hinted at a mass loss mechanism that was $Z$ independent, but now \citet{Higgins2020} revealed a key indirect $Z$-dependent wind effect: at low $Z$, strong (semiconvective) mixing kept the HD limit low, while stronger winds at higher $Z$ prevented semiconvective regions from forming, naturally leading (without tweaks) to a higher Galactic HD limit of order $\log (L/L_\odot) \approx 5.8$.

More recently, observational results of \citet{Davies2020} suggest the Galactic HD limit also to be as low as $\log (L/L_\odot) \approx 5.5$, and if this finding is confirmed this would imply that the physics setting the HD limit remains elusive. For this reason, we here critically investigate the physics of superadiabaticity in radiative layers close to the Eddington limit utilising the stellar evolution code MESA. While the temperature independent part of the HD limit can extend to cooler BSGs with $\log (T_{\text{eff}}/\mathrm{K}) \approx 4$, in this study we primarily focus on reproducing the Z-independent limit for the red supergiants, the RSG luminosity limit.

The 1D hydrostatic models of massive stars close to their Eddington limit predict the occurrence of very low density, highly inflated envelopes for both core-hydrogen burning \citep{Sanyal2015} as well as pure He stars \citep{Graf2012}. In reality, the different processes that capture the microphysics inside the envelopes of these stars are 3 dimensional and would ideally be modelled by 3D hydrodynamic codes. Efforts in the 3D modelling of stellar envelopes might suggest a suppression of inflation effects \citep[e.g.][]{Jiang2015}, but additional insight into 3D mixing processes of stellar envelopes is still desirable. 

The current situation of the modelling of stellar envelopes in state-of-the-art stellar structure and evolution models is mixed. The BEC ("Bonn") models \citep{Koehler2015, Sanyal2015} solve the issue of super-Eddington near-surface layers and associated density inversions by allowing the envelopes to inflate to large radii, while the GENEC ("Geneva") models \citep{Yusof2013} effectively solve the issue by swapping the pressure scale height for the density scale height. The oft-used MESA suite allow the problem to be resolved through an alternate energy transport mechanism that induces `excess' mixing in the stellar envelope within a routine dubbed MLT++ \citep{MESA13}. 

The resolution to these model uncertainties may ultimately be found in 3D radiation hydrodynamical modelling, but all current hydrodynamical models still need to make the necessary approximations with respect to the 3D radiative force in clumped and porous media \citep[e.g.][]{Shaviv2000, Owocki2015}.
Alternatively, progress might be made through the comparison of stellar models with the observed upper-luminosity limit as a function of metallicity $Z$, as explored in this paper.

The paper is structured as follows. In Sect. \ref{sec:methods}, we discuss the different input physics used in the 1D modelling of massive stars including mass loss, different mixing processes and rotation. In Sect. \ref{sec:results}, we discuss the evolution of massive stars in the Hertzsprung-Russell diagram (HRD) and investigate the effects of mass loss and different mixing conditions during both the main sequence and post-main sequence evolution of RSGs. We also discuss the upper luminosity cutoff of the coolest supergiants and how different processes affect this limit as a function of metallicity. Finally, the discussions are presented in Sect. \ref{sec:disc} and concluded in Sect. \ref{sec:conc}. 

\section{Method}
\label{sec:methods}
\subsection{MESA models}

The one-dimensional stellar evolution code Modules for Experiments in Stellar Astrophysics (MESA version 12115) \citep{MESA11, MESA13, MESA15} is used to compute our grid of stellar  models. The models are used to investigate the cutoff luminosity of red supergiants as a function of metallicity. We begin by highlighting the effects of excess envelope mixing by comparing models with and without the treatment of superadiabatic mixing (see Sect. \ref{sec:HRD}). This is followed by grids of massive star models consisting of masses from $15$ $M_{\odot}$ to $37.5$ $M_{\odot}$ at intervals of $2.5$  $M_{\odot}$, at three different initial metallicities and two different values of overshooting mixing above the hydrogen core - low ($f_{\text{ov}} = 0.01$) and high ($f_{\text{ov}} = 0.05$)\footnote{see Appendix \ref{appendix:C} for various overshooting options in MESA and note that an exponential overshooting parameter $f_{\mathrm{ov}}$ corresponds to a step overshooting parameter $\alpha_{\mathrm{ov}}$ by an approximate relation $f_{\text{ov}} \approx \alpha_{\text{ov}}/10$}, a total of 10$\times$3$\times$2 =  60 initial models. Further models are run to test the effects of certain processes such as mass loss, rotation and overshooting during core helium burning on our results. All the models are evolved until core helium exhaustion when the central helium mass fraction $(Y_c)$ drops below 0.01. 

All models start with a uniform composition with a heavy metal fraction scaled with the solar abundance values provided by \citet{GS98}. The three sets of models for the Galactic, LMC and SMC metallicities have $Z_{\text{GAL}} = 0.017$, $Z_{\text{LMC}} = 0.008$ and $Z_{\text{SMC}} = 0.004$ respectively. The initial helium mass fraction in our models are evaluated as follows: $Y = Y_{\text{prim}} + (\Delta Y/\Delta Z) Z$ using MESA default values of $Y_{\text{prim}} = 0.24$ and $(\Delta Y/\Delta Z) = 2$. 

Convection is treated using the standard mixing length theory developed by \citet{MLT68} with the free parameter $\alpha_{\text{MLT}} = 1.82$\footnote{We also test for a lower value of $\alpha_{\text{MLT}}=1.5$ like \citet{Higgins2019}. We also note that $\alpha_{\text{MLT}}$ can change with mass and metallicity, for example \citet{Bonaca2012, Viani2018, Song2020}, which we do not explore further in this paper.} \citep{CHOI16}. The exact value of $\alpha_{\text{MLT}}$ can change the RSG temperatures by a few hundred Kelvin, but has negligible effects on the stars that evolve bluewards and on the RSG limit. Overshooting mixing above or below all the convective regions uses a diffusive approach with an exponential profile as described in \citet{Herwig2000}. All core hydrogen burning models have overshooting regions above their convective core, with high or low efficiencies as mentioned above\footnote{We also include the effects of overshooting above the convective shells with $f_{\text{ov}}$ value equivalent to that above the hydrogen burning core.}. 
The overshooting above the convective core during core He burning is set to $f_{\text{ov}}$ = 0.01 (but see Sect. \ref{sec:he_os}). We use the Ledoux criteria for convection, with a high semiconvective efficiency of $\alpha_{\text{sc}} = 100$. A thermohaline mixing coefficient of $\alpha_{\text{th}} = 1$ is used.

We also use the predictive mixing routine, introduced in \citet{MESA18}, that carefully deals with the definition of convective boundaries, allowing the boundaries to satisfy the equality of radiative and adiabatic temperature gradient on its convective side. This scheme was further improved in the subsequent paper as convective premixing (CPM), that replaced the `problematic' convective regions (where $\nabla_{\text{rad}} \neq \nabla_{\text{ad}}$ on the convective side of the boundary) with adiabatically stratified semiconvective regions. The presence of such semiconvective regions, was first discussed in the seminal paper by \citet{SH1958}. Implementation and consequences of this scheme are  discussed in \citet{MESA19}

\subsection{Mass loss}

Mass loss in our models is implemented similar to the "Dutch wind scheme" comprising of three different recipes as outlined below. Hot O stars ($T_{\text{eff}}>10\,$kK) with surface hydrogen mass fraction $X_s>0.4$ use the metallicity-dependent mass loss recipe from \citet{Vink2001}. The mass-loss rates used in our models scale with the surface iron abundance ($Z_{\text{Fe}}$) instead of the total metal fraction at the surface. This scheme also features the so-called first bistability jump at $\log (T_{\text{eff}}/\mathrm{K})\approx 4.4$. The sudden changes in the ionization fraction of iron at this temperature, namely recombination of Fe\,\textsc{iv} to Fe\,\textsc{iii}, results in mass-loss rates on the cool side of the jump almost five times higher compared to the hot side. The nature and origin of such bi-stable winds is discussed in detail in \citet{Vink1999}. MESA implements this jump by interpolating between the hot and cool side, with the jump temperature scaling with the host metal mass fraction \textit{Z}$_{\text{init}}$ \citep{Vink2001}. The second bistability jump at $T_{\text{eff}}\lesssim 12,500\,$K due to recombination of Fe\,\textsc{iii} to Fe\,\textsc{ii}, is currently not implemented in the standard wind prescription of massive stars in MESA. As there is no evidence for a strong second bistability jump at sub-solar metallicities \citep[see][Fig. 1]{Vink2001}, we do not implement this jump in our models.

Cooler stars with $T_{\text{eff}}<10\,\text{kK}$ use the dust driven mass loss recipe from \citet{deJager1988}. For temperatures $10\,\text{kK}<T_{\text{eff}}<11\,\text{kK}$, we interpolate between the rates derived from \citet{Vink2001} and \citet{deJager1988}. Post-main sequence stars that lose their envelope ($X_s<0.4$) and evolve bluewards use the recipe adopted from \citet{NL2000}. As seen from Fig. \ref{fig:HD1}, a very small number of our models (specifically at Galactic metallicity with higher overshooting) reach the Wolf-Rayet phase and use the aforementioned WR recipe, and thus have no effect in setting the HD limit. The setup for testing the effect of mass-loss rates, especially during post-main sequence, is discussed below.

Apart from this scheme, we have also tested the effect of having radiation-driven winds instead of dust-driven winds in the cool supergiant regime on the location of the RSG luminosity limit. We test this by replacing the metallicity independent de Jager mass-loss rates in the yellow supergiant regime at temperatures between $4000\,\text{K}<T_{\text{eff}}<10\,$kK with a mass loss that scales with the surface iron abundance, similar to the mass-loss rates used in \citet{Vink2020}. This is done by extrapolating the mass-loss rates from the cool side of the Vink wind (which is lower by an order of magnitude) all the way down to $5500\,$K where it transitions to the de Jager rates. This switch from a metallicity-dependent to a metallicity-independent wind effectively lowers the mass-loss rates at these temperatures. For the metallicities considered in our study, such a switch in mass-loss rates would affect the SMC models the most. The comparison of the two scenarios and implications of having lower mass-loss rates are discussed in Sect. \ref{section:mass_loss}.

\subsection{Mixing processes}
\label{section:mixing}

The luminosities of massive stars increase steeply with initial stellar mass resulting in large $L/M$ ratios that bring these stars closer to their Eddington limit. The local, radius-dependent Eddington parameter $\Gamma(r)$ inside a star at a mass coordinate $m(r)$ with local luminosity $L(r)$ is given by
\begin{equation}
\begin{array}{c@{\qquad}c}
\Gamma(r) = \dfrac{\chi(r)L(r)}{4\pi Gcm} > \Gamma_\text{e}
\end{array}
\label{eq:relate_beta_lambda}
\end{equation}
where $\chi(r)$ is the total opacity at radius $r$ that takes into account the opacity contributions from both electron scattering as well as metal lines. The Eddington limit is defined by the condition when the radiative acceleration (considering all the opacity contributions) pointing outward balances the inward gravitational acceleration. The above equation takes into account the energy transported by both radiation as well as convective motions. Subtracting the energy transported by convection we get
\begin{equation}
\begin{array}{c@{\qquad}c}
\lambda(r) =  \Gamma_{\text{rad}}(r)  = \Gamma(r) - \Gamma_{\text{conv}}(r) =  \dfrac{\chi(r)L_{\text{rad}}(r)}{4\pi Gcm} 
\end{array}
\label{eq:relate_beta_lambda}
\end{equation}
The values of $\lambda(r)$ in the convective core of massive stars remain well below the Eddington limit, even when $\Gamma(r)$ increases beyond unity. However, this is not true closer to the stellar surface, where layers near the ionization zones of H, He and Fe can have significantly higher values of $\lambda(r)$. As the Eddington parameter $\lambda(r)$ approaches unity near one of these ionization zones, the stars are predicted to develop very low density envelopes inflated to very large radii. Owing to very low densities in these envelopes, radiation pressure begins to dominate. The contribution of radiation pressure to the total pressure is quantified by $\beta(r)$ defined as
\begin{equation}
\begin{array}{c@{\qquad}c}
\beta(r) = \dfrac{P_{\text{gas}}}{P_{\text{tot}}}, \;\; 1-\beta(r) = \dfrac{P_{\text{rad}}}{P_{\text{tot}}}
\end{array}
\label{eq:relate_beta_lambda}
\end{equation}
These inflated envelopes are also characterized by a density inversion close to the surface, while maintaining hydrostatic equilibrium. This predicted \textit{inflation} effect occurs when the stars are close to their Eddington limit, and is different from the radial \textit{expansion} of hydrogen shell burning stars that quickly traverse through the Hertzsprung Gap phase. During the main sequence, $\lambda(r)$ values of $\gtrsim$ 0.9 are reached for stars of initial masses $\gtrsim 40 \,M_\odot$. Super-Eddington conditions are easily reached for core helium burning stars that have access to opacity bumps located at cooler temperatures, namely the H/He\,\textsc{i} and He\,\textsc{ii} bumps. For example, cooler supergiants can have $\lambda(r)$ reaching values as high as $\approx 7$ \citep{Sanyal2015}, corresponding to the ionization zone of hydrogen. 

The inflated, low density envelopes, especially during phases involving cooler supergiants, make energy transport by convection highly inefficient, resulting in superadiabatic temperature gradients. The degree of superadiabaticity inside the star is defined as follows
\begin{equation}
\begin{array}{c@{\qquad}c}
x_{\text{sa}} = \nabla_{T}  - \nabla_{\text{ad}}, 
\end{array}
\label{eq:relate_beta_lambda}
\end{equation}
where $\nabla_{T}$ and $\nabla_{\text{ad}}$ are the actual and adiabatic temperature gradients inside the star. The superadiabaticity gives a measure of the inefficiency of convection. Layers inside the star with $x_{\text{sa}} < 0$ are radiative regions, while $x_{\text{sa}} > 0$ denotes convective regions. For efficient convective regions deep inside the star, the temperature gradient follows the adiabatic temperature gradient ($\nabla_{T} \rightarrow \nabla_{\text{ad}}$) giving a degree of superadiabaticity very close to zero but positive. Inefficient convective zones that form closer to the surface of the star have higher values of superadiabaticities, with the degree of superadiabaticity increasing with more inefficient convection. For our study, regions with $x_{\text{sa}}  \gtrsim 10^{-4}$  (see the parameters used in Appendix \ref{appendix:A}) are considered superadiabatic regions.

Modelling such inflated envelopes with density inversions at the surface (and the treatment of superadiabatic convection) not only pose numerical challenges, but are sometimes regarded as unphysical, and are usually removed by different code specific routines. The absence of cooler supergiants above $\log (L/L_\odot) \approx 5.5$ also hints at an alternate mixing mechanism that drives stars towards the blue as they approach their Eddington limit. Thus stellar evolution codes such as MESA implement a mixing routine that is based on the closeness of the model to its Eddington limit and the predominance of radiation pressure.

As mentioned above, the proximity of the star to its Eddington limit resulting in inflation is sometimes considered unphysical. The alternative, that is investigated in this paper, is the effect of excess envelope mixing as the star approaches its Eddington limit. This excess mixing occurs in the radiation dominated inefficient convective layers, that prevents the inflation of the star and erases the density inversions near the surface. Instead the excess mixing keeps the star compact causing it to evolve bluewards, preventing the formation of cooler supergiants above a certain luminosity limit. How this mixing affects the upper luminosity limit of RSGs as a function of their initial metallicity is discussed in Sect. \ref{sec:HD_limit}. Moreover, the modelling of highly inefficient convective envelopes with density inversions 
poses numerical challenges in numerical codes such as MESA. These envelopes have very short thermal timescales (comparable to its dynamical timescale) causing timestep problems in numerical implementations.

Both the issues of having excess mixing instead of allowing the stars to inflate when close to their Eddington limit and the numerical challenges they pose for reliable modelling is tackled by the MLT++ routine in MESA. We use this routine as a proxy to simulate excess mixing when stars are dominated by radiation pressure and close to their Eddington limit. It effectively suppresses the inflation of massive stars during their RSG phase, as well as during main sequence although to a lesser extent, and allows the models to get rid of their envelopes without numerical issues. The standard set of MLT++ parameters is used in our models. The amount by which the superadiabaticity is reduced is governed by an efficiency factor $\alpha$, with $\alpha = 0$ meaning no reduction in superadiabaticity, $\alpha = 1$ meaning the superadiabatic layers receive a full boost in efficiency and values in between receiving a partial boost in the efficiency. The value of $\alpha$ is primarily decided by the values of the maximum $\lambda(r)$ and the minimum $\beta(r)$ inside the star. The complete parameter space of MLT++ and the conditions to enable it is provided in Appendix \ref{appendix:A}. The effect of MLT++ on the location of the extrema of $\lambda(r)$ and $\beta(r)$ as the star evolves is discussed in Appendix \ref{appendix:B}.

\subsection{Rotation}

Here we discuss the possible effects of stellar rotation on the luminosity cutoff of red supergiants. Two sets of rotating models are computed that have equatorial velocities of 0.3 and 0.4 times the critical velocity ($\Omega/\Omega_{\text{crit}} = 0.3\; \text{and} \;0.4$) at ZAMS. Implementation of mixing and angular momentum transport in the star due to rotational instabilities and their corresponding factors follows \citet{Heger2000}. The effects of rotational induced enhanced  mass loss is not considered in this work \citep[see][]{Muller2014, Higgins2019}. With this setup for rotating stars, we run models at all three metallicities until stars no longer evolve to the RSG phase. We find that rotating models are more luminous compared to their non-rotating counterparts, as they have a larger core size owing to rotation induced mixing. One would expect this to change the luminosity limit of the supergiants. However, rotational mixing also results in hotter evolution both during the main sequence and post-hydrogen exhaustion, causing slow and moderately rotating stars to move upwards and bluewards in the HRD. This is the case for stars rotating with velocities slower than $\Omega/\Omega_{\text{crit}} \approx 0.4$. For even faster rotators with ($\Omega/\Omega_{\text{crit}} \gtrsim 0.4-0.6$), they are predicted to undergo chemical homogeneous evolution, especially at lower metallicities, quickly evolving towards hotter temperatures even during the main sequence \citep{Maeder1987, YL2005, WH2006, Brott2011} . These fast rotators do not form supergiants at any mass range and thus are not responsible for setting the RSG limit. The luminosity limit of supergiants remains almost unchanged, with the slow rotators shifting upwards and bluewards, while the fastest rotators undergo chemical homogeneous evolution and do not influence the cutoff luminosity. Therefore we investigate the case of non-rotating models in the discussion below, however the results apply for rotating stars as well.

\section{Results}
\label{sec:results}

\subsection{Stellar tracks and the HRDs}
\label{sec:HRD}

\begin{figure*}
    \includegraphics[width = \textwidth]{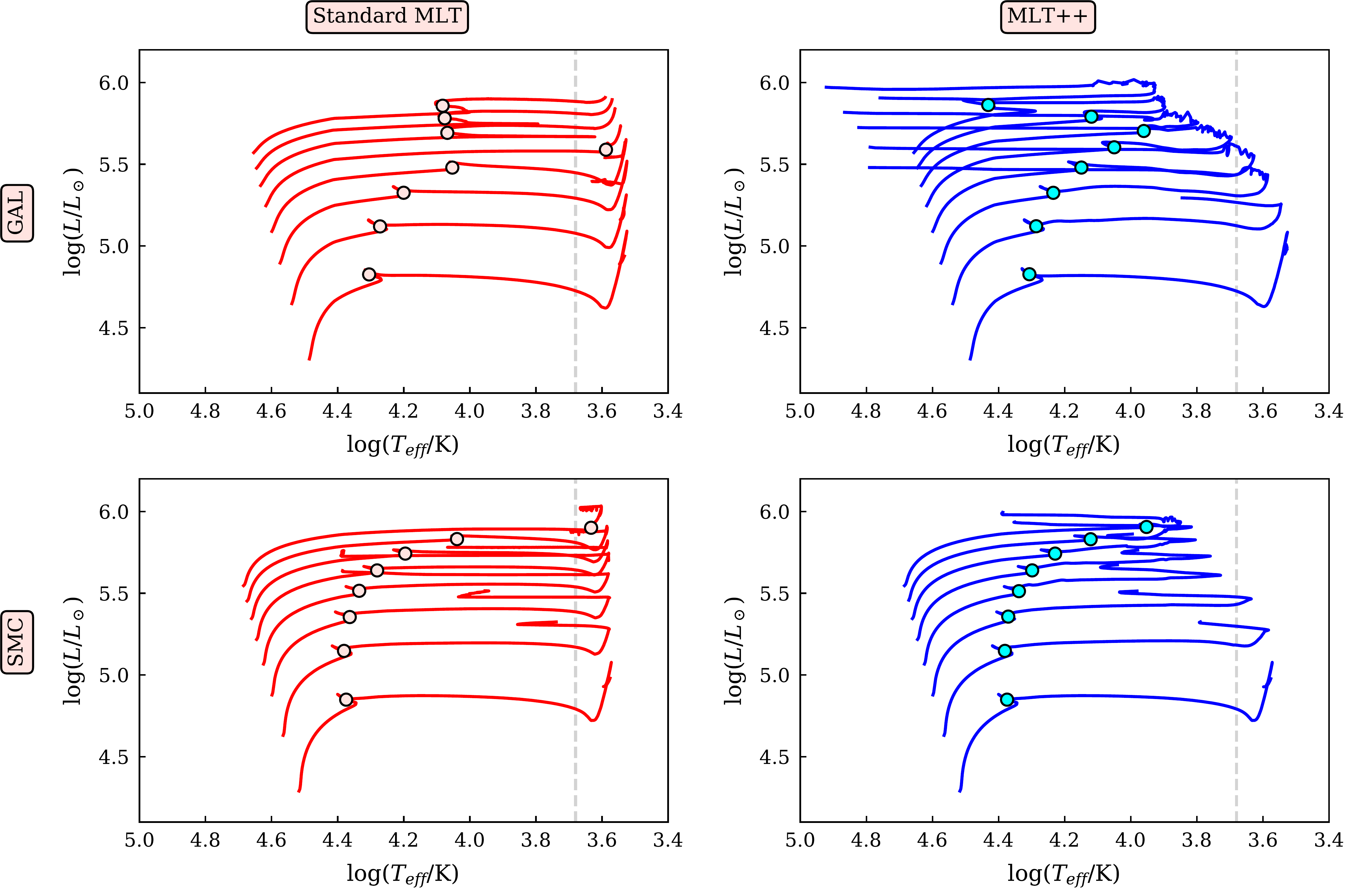}
    \caption{HRD evolution of massive star models with initial masses ranging from 15 to 50 $M_\odot$ highlighting the effects of envelope mixing on the location of TAMS and post-main sequence evolution. Models on the left (red) are computed allowing models to fully inflate, while the models on the right (blue) include excess envelope mixing, resulting in a blueward evolution. The terminal age main sequence is represented by circles. The models are computed for an overshooting value of $f_{\text{ov}} = 0.05$, for the two extreme initial metallicities considered in this study: Galactic (top) and SMC (bottom). In order to test a Z-independent RSG limit, we investigate the effects of envelope mixing with high overshooting implemented in both sets of models. } 
    \label{fig:with_without_MLT++}
\end{figure*}

\begin{figure*}
    \includegraphics[width = \textwidth]{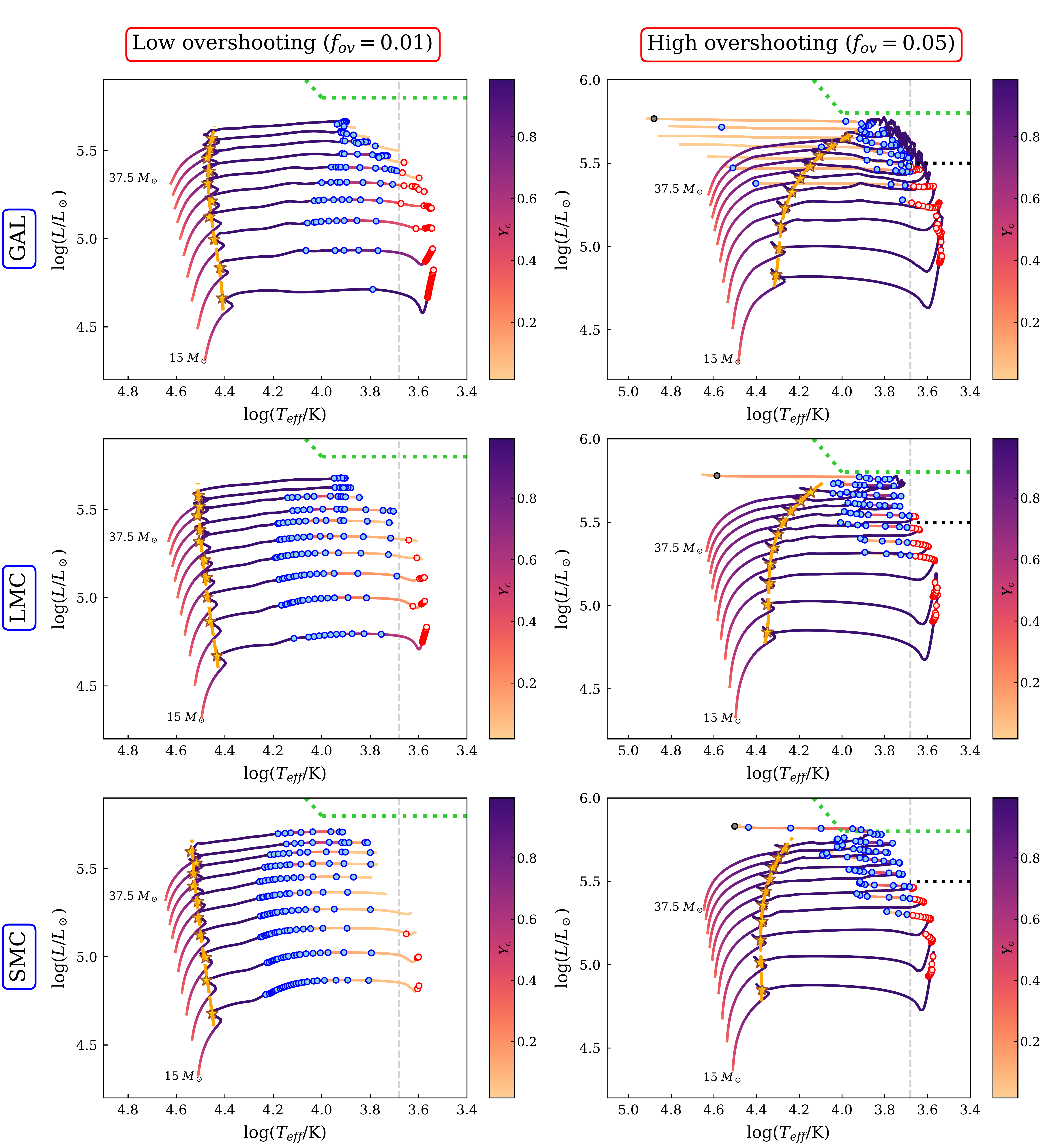}
    \caption{Evolutionary tracks of single star models of initial mass $15\,M_\odot$ to $37.5\,M_\odot$ computed at three different initial metallicities: Galactic(\textit{top}), LMC (\textit{middle}) and SMC (\textit{bottom}) having two different overshooting efficiencies: low (\textit{left}) and high (\textit{right}). The plots show HRD of our models starting from the zero-age main sequence until end of He burning ($Y_c < 0.01$). The colorbar represents the central helium mass fraction. The yellow stars mark the exhaustion of hydrogen in the core and the subsequent dots are marked every 50,000 years during core He burning. Blue circles represent the blue supergiant regime (with $\log (T_{\text{eff}}/\mathrm{K}) > 3.68$) and red circles represent evolution of the star in the red supergiant regime (with $\log (T_{\text{eff}}/\mathrm{K}) < 3.68$). The black circles represent those models where the surface hydrogen fraction has fallen below 0.4, thus triggering the WR mass loss recipe. The orange dashed line marks the location of the TAMS with increasing mass. The gray dashed line at $\log (T_{\text{eff}}/\mathrm{K}) = 3.68$ separates the two regions. The green dotted line marks the classical empirical HD limit with the temperature independent section located at $\log L/L_\odot = 5.8$. The black dotted line marks the down-revised empirical limit for the RSGs at $\log L/L_\odot = 5.5$}
    \label{fig:HRD}
\end{figure*}

\begin{figure*}
    \includegraphics[width = \textwidth]{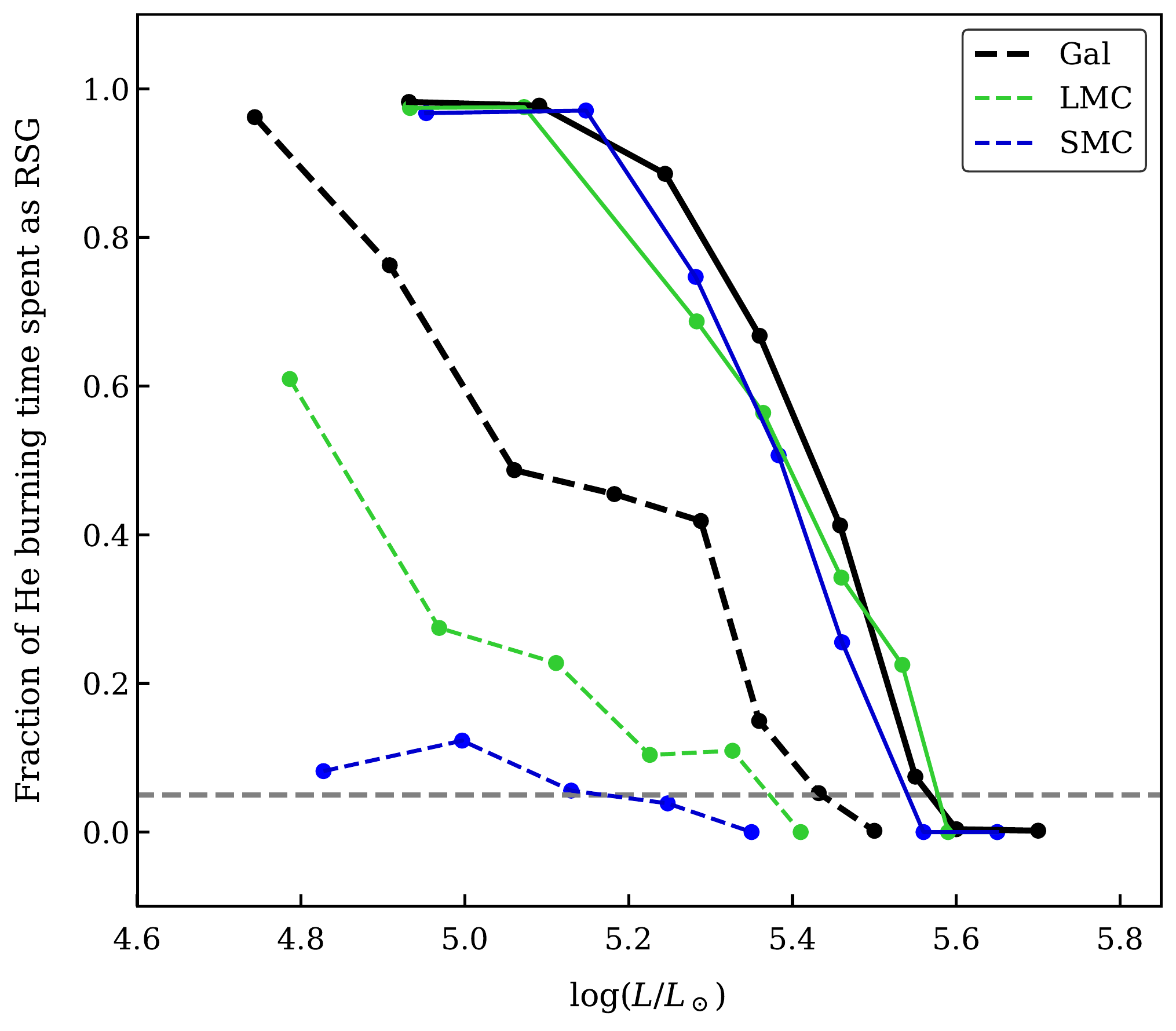}
    \caption{Distribution of fraction of He burning time (in Myr) spent below the threshold effective temperature $\log (T_{\text{eff}}/\mathrm{K}) = 3.68$ as a function of the luminosity of the RSG, obtained by taking the average luminosity of the red dots in Fig. 1 corresponding to one single evolution track . The dashed color lines correspond to low overshooting and solid color lines denote high overshooting. The black, green and blue correspond to Galactic, LMC and SMC-like metallicity respectively. The grey dashed line marks the 5$\%$ limit that we assume to set the HD limit.}
    \label{fig:HD1}
\end{figure*}


\citet{HD1979} showed that the cooler supergiants in the Galaxy as well as LMC, with $T_{\text{eff}} < 15\,$kK, have an upper luminosity cutoff  that occurs near bolometric magnitude of $M_{\text{bol}} \approx -9.5$ to $10$ mag. Taking the bolometric magnitude of the Sun to be $-4.75$ mag, gives the temperature independent part of the HD limit to occur at $\log (L/L_\odot) \approx 5.8$. But \citet{Davies2018} recently studied the luminosity distribution in these galaxies and found lower limits of $\log (L/L_\odot) \approx 5.5$ for both metallicities, and surprisingly a slightly lower limit at SMC metallicity compared to the LMC. A majority of the RSGs observed in \citet{Davies2018} have luminosities of $\log (L/L_\odot) \approx 5.4-5.5$ or less. However, there is one object in each of the Magellanic Clouds observed above the limit. Both \citet{Davies2018} and \citet{Higgins2020} hint at the fact that the RSG limit is not a "hard" border where stars entirely avoid the region above it. RSGs can evolve above the limit, but spend such small fraction of their helium burning time above the limit that they are rarely observed. Following \citet{Davies2020}, a further downward revision of the luminosities in the Galaxy with a lack of cooler supergiants with $\log (L/L_\odot) > 5.5$ hints at a Z-independent HD limit. This includes the population of RSGs, whose empirical upper luminosity limit we try to reproduce using excess mixing in their superadiabatic envelopes.

As described in Sect. \ref{section:mixing}, we investigate the role of excess envelope mixing, choosing to suppress the effects of inflation. For this purpose, we show the HRD evolution of massive star models with (blue) and without (red) the treatment of superadiabatic mixing in Fig. \ref{fig:with_without_MLT++}. The models shown here are computed for overshooting efficiency of $f_{\text{ov}} = 0.05$. The figure demonstrates the following two key points: (1) models which include extra envelope mixing  above a certain luminosity no longer evolve as RSGs, but instead turn bluewards in the HRD. This lowers the maximum luminosity of helium burning stars evolving as RSGs, i.e., it lowers the RSG luminosity limit, and (2) the implementation of excess envelope mixing can essentially lower the RSG limit even at high overshooting. We elaborate further on this below.

The effects of mixing in the envelope of massive stars are already visible during the main sequence. Our higher overshooting models that allow for inflation evolve towards very cool temperatures, some even cooler than the threshold temperature of $\log (T_{\text{eff}}/\mathrm{K}) = 3.68$ we have assumed for the definition of RSGs (see below). The stars that dip into the RSG regime already during the MS phase above logL = 5.5, their timescales should be long enough to be observed at such high luminosities. This behavior is also observed regardless of the initial metallicity. In comparison, models that efficiently mix the sub-surface superadiabatic layers have a significantly hotter TAMS and smaller radii, as the effects of inflation are suppressed.

The post-main sequence evolution of our models are also substantially affected by this mixing mechanism. High overshooting models without MLT++ begin core helium burning as RSGs at luminosities as high as $\log (L/L_\odot) \approx 6$. The highest mass models on the left do not finish burning helium due to numerical issues as mentioned in Sect. \ref{section:mixing}. However, in the mass range which sets the HD limit ($\sim 30-40 \;\text{M}_\odot$) we can probe the evolution of masses up to $30-35 \; \text{M}_\odot$ for the Galactic models for the majority of He-burning during the RSG phase, while at SMC this is increased even to $40-50 \; \text{M}_\odot$ models. These models spend a considerable fraction of their helium burning time above the current empirical estimates of the limit of $\log (L/L_\odot) \approx 5.5$. In comparison, models with excess envelope mixing evolve bluewards as post-RSG blue or yellow supergiants above a certain limit. Thus the treatment of superadiabatic convection can potentially lower the RSG limit.

A recent study on the HD limit by \citet{Higgins2020}, explored the effects of semiconvective mixing and the indirect effect of mass loss in setting this limit. They were able to reproduce the observed RSG luminosity limit for LMC and SMC metallicities of $\log (L/L_\odot) \approx 5.5$, with the downward revision of RSG luminosities in the Magellanic Clouds as reported by \citep{Davies2018}. But this required models with low overshooting efficiency of $\alpha_{\text{ov}} = 0.1$ (corresponding to $f_{\text{ov}}\approx0.01$), that allowed the formation of semiconvective layers above the convective core during the main sequence. The semiconvective mixing kept the stars bluer resulting in the lowering of the RSG limit. The limit at galactic metallicity remained at $\log (L/L_\odot) \approx 5.8$, in agreement with the then observed galactic limit, as higher mass-loss rates prevented the formation of these semiconvective regions.

Semiconvection as a mixing mechanism can cause stars to evolve bluewards effectively lowering the luminosity limit, as described above, as long as semiconvective layers can actually form above the hydrogen core. Larger mixing regions above the core, or higher mass-loss rates at higher initial metallicity can suppress these layers, thus enabling semiconvection to take effect only at lower metallicities (LMC and SMC) and for low overshooting efficiency ($\alpha_{\text{ov}} = 0.1$). However, various studies have invoked higher overshooting efficiencies near the convective boundary to explain observations \citep{Vink2010, Castro2014, Higgins2019}. Excess envelope mixing during post-main sequence offers an alternative mechanism that enables the blueward evolution of massive stars even at high overshooting. Here we conduct a complementary study by thoroughly investigating the effects of excess mixing (instead of inflation) in setting the RSG limit as a function of metallicity. 

We present our grid of massive star models which implement the treatment of superadiabatic mixing in stars close to their Eddington limit, while also allowing for higher values of overshooting up to $f_{\text{ov}}\approx0.05$. In Fig. \ref{fig:HRD} we show six HRDs with the post-main sequence evolution marked by dots every 50,000 years. We consider RSGs as supergiants cooler than a threshold surface temperature $\log (T_{\text{eff}}/\mathrm{K}) = 3.68$ following \citet{Drout2009}. We assume the location of the luminosity limit of RSGs $L_{\text{RSG}}$ from our models such that above this limit stars spend less than 5$\%$ of their helium burning time below $\log (T_{\text{eff}}/\mathrm{K}) = 3.68$. Such a definition is adopted keeping in mind that the RSG limit is not a "hard" border, and stars can spend a small fraction of their helium burning time above the limit. 

\subsection{Overshooting during main sequence}
\label{sec:OS_H}

The three HRDs on the left in Fig. \ref{fig:HRD} show the evolution of stars with low overshooting above their cores, whereas the HRDs on the right employ higher overshooting during main sequence. The location of the TAMS (gold stars) in the right column with higher overshooting is shifted to higher luminosities and cooler temperatures (orange dashed line), due to a larger fuel supply during the main sequence. The amount of mixing near the boundary of convective regions directly affects the effective size of the core during hydrogen burning. Moreover, this also has a significant effect on the post-main sequence evolution of our models, namely the evolution of the stellar radius during helium burning. The models with lower overshooting tend to begin their helium-burning as blue supergiants ($\sim 100\,R_\odot$) and spend a considerable fraction of their helium burning phase as pre-RSG objects (at all metallicities) as evident by the cluster of blue dots at temperatures greater than $\log (T_{\text{eff}}/\mathrm{K}) = 3.68$. Also evident from the left column is the effect of metallicity on the radial expansion of the star during helium burning, with the stars remaining relatively compact for lower metallicity. \citet{Klencki2020} recently discussed the effect of metallicity on the radial expansion of stars in the context of binary interaction, with stars at lower metallicity remaining compact during the post-main sequence. With a strong effect of metallicity on the radial expansion of our models at low overshooting combined with them beginning their helium-burning phase as a BSG, these stars show a clear metallicity-dependent RSG limit (also see dashed lines in Fig. \ref{fig:HD1}).

\begin{figure}
    \includegraphics[width = \columnwidth]{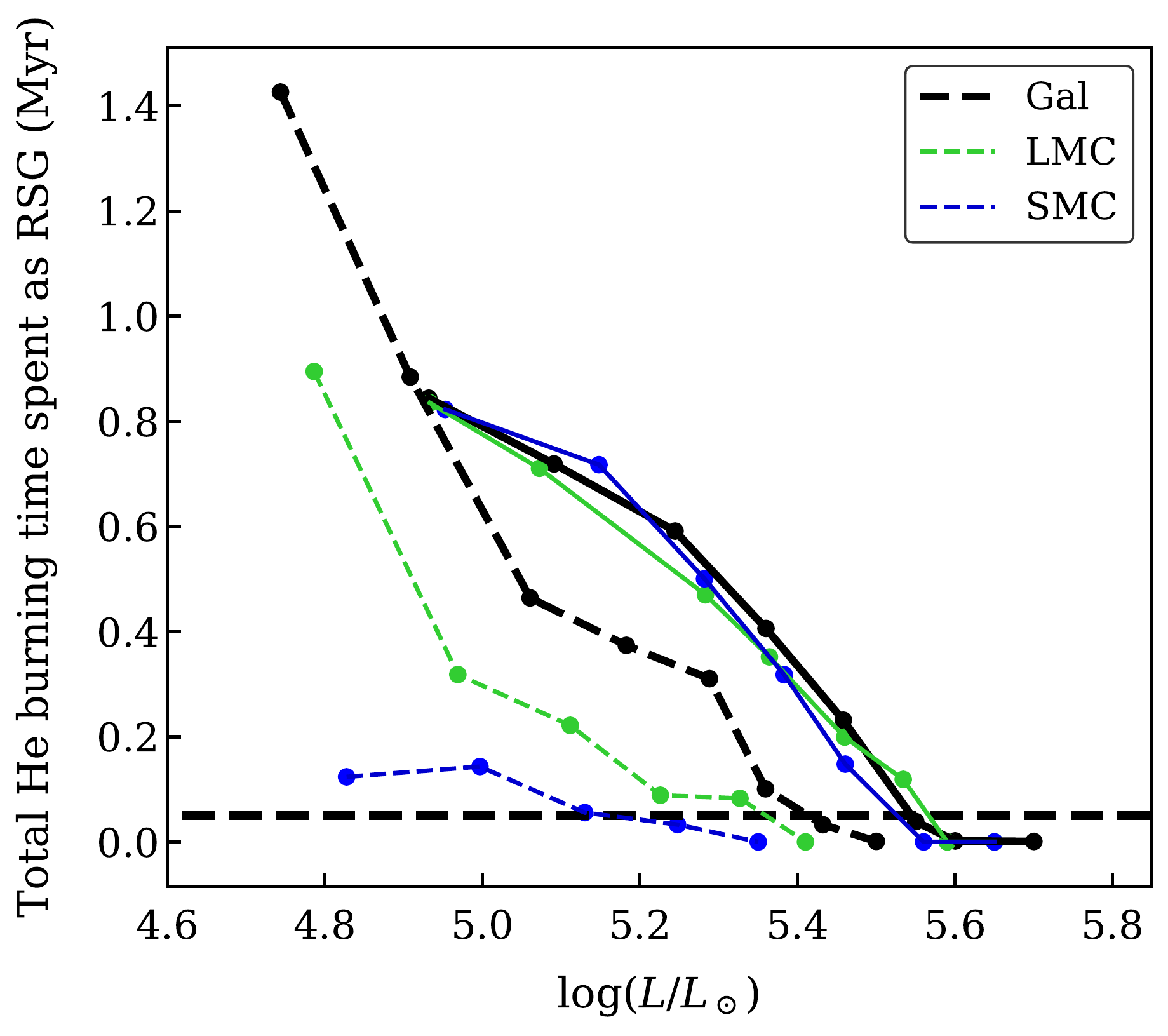}
    \caption{Distribution of total He burning time spent below the threshold effective temperature $\log (T_{\text{eff}}/\mathrm{K}) = 3.68$ as a function of the luminosity of the RSG, obtained as discussed previously. The dashed and solid color lines have the same meaning as Fig. \ref{fig:HD1}. The grey dashed line marks a limit on the total time spent as an RSG, below which stars spend less than 50,000 years in the RSG regime.}
    \label{fig:HD2}
\end{figure}

\begin{figure}
    \includegraphics[width = \columnwidth]{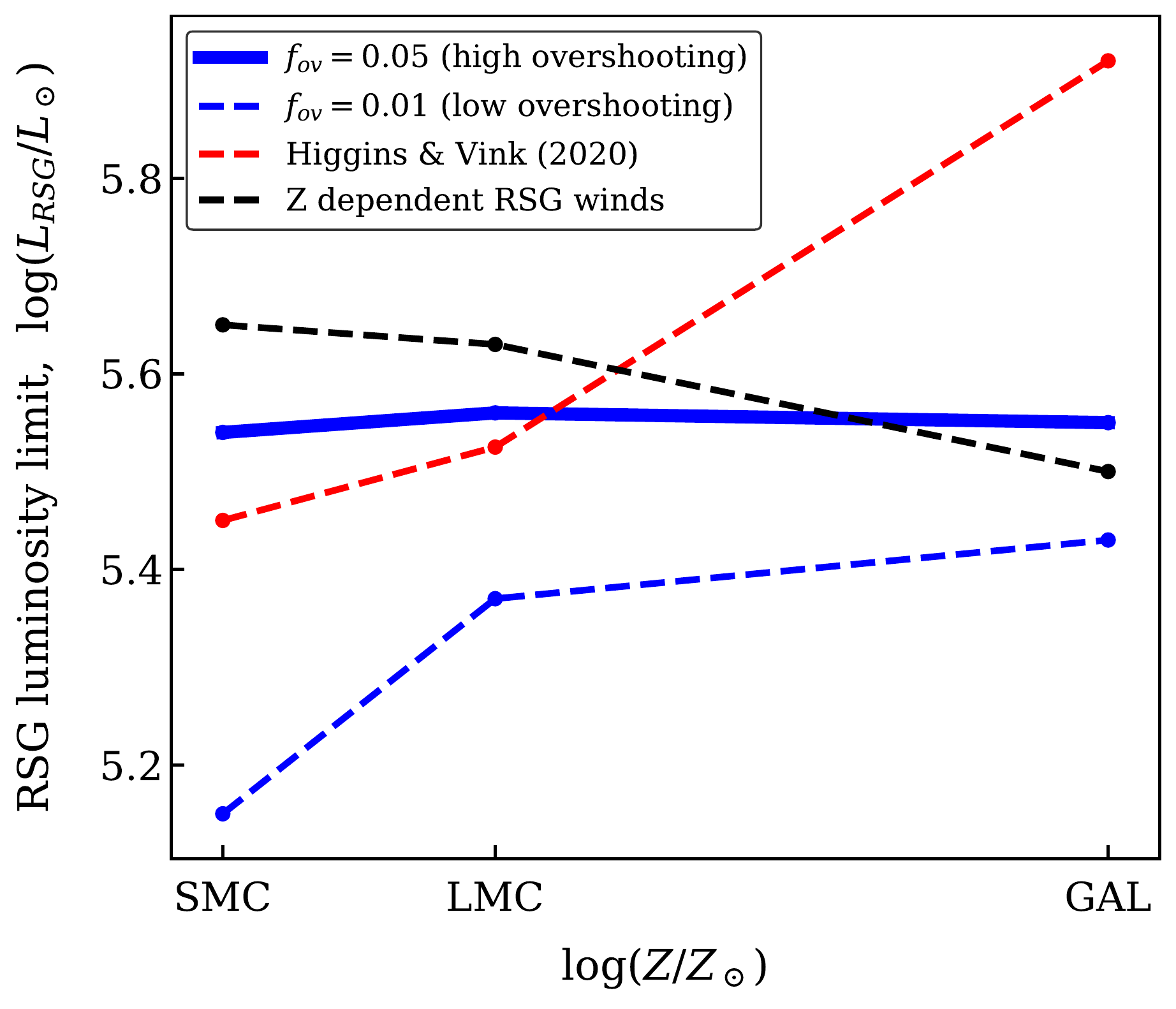}
    \caption{The upper luminosity limit of RSGs as inferred from our models, with the blue dashed and solid lines representing models on the left and right of Fig. \ref{fig:HRD} with overshooting parameters $f_{\text{ov}}$ = 0.01 and 0.05 respectively. The limit as represented by the red dashed line is obtained from \citet{Higgins2020} (see their Fig. 6). The black dashed line denotes models with lowered mass-loss rates assuming metallicity-dependent mass loss for all temperatures. } 
    \label{fig:low_mass_loss}
\end{figure}

However, the models with higher overshooting begin their helium burning either as an RSG or during blueward evolution without entering the RSG regime, regardless of the metallicity, as seen from the absence of blue dots in the Hertzsprung-gap phase for the HRDs on the right. The very similar post-main sequence evolution of stars that begin their helium burning as RSGs regardless of their metallicity is important in understanding why the excess mixing in their superadiabatic layers is triggered at approximately the same luminosity across all three metallicities. With our definition of the location of the luminosity limit of RSGs as the luminosity above which stars spend less than $5\%$ of their helium burning time as RSGs, all models with higher overshooting have a cutoff luminosity of $\log (L/L_\odot) \approx 5.5$, in agreement with the current observational constraints on the luminosities of RSGs (black dotted line). This is a factor of two lower compared to the observed limit at $\log (L/L_\odot) \approx 5.8$ (green dotted line) as originally found in \citet{HD1979}. Fig.\,\ref{fig:HD1} shows the distribution of percentage time spent in the RSG regime by our models as a function of their luminosity, with the grey dashed line marking the 5$\%$ limit. Models with higher overshooting during the main sequence are equally affected by the superadiabatic treatment of the envelope, and all have a very similar luminosity limit of $\log (L/L_\odot) \approx 5.5$ regardless of metallicity (see also blue solid line in Fig. \ref{fig:low_mass_loss}). The Z-independent RSG limit is set by the high overshooting models. As discussed previously the effect of semiconvection is negligible at higher values of overshooting, thus our results do no depend on the semiconvective efficiency or whether Schwarzschild or Ledoux convection criteria is used.

This behavior is also observed when the total helium burning time spent as an RSG is plotted as a function of luminosity (Fig. \ref{fig:HD2}). Comparing the three solid lines in the two figures, the post-main sequence evolution across all three metallicities is very similar. This metallicity independent behavior of of the RSG luminosity limit is not observed for lower overshooting, where the limits shift to even lower values with the SMC-like metallicity having a cutoff of $\log (L/L_\odot) \approx 5.1$ (see also blue dashed line in Fig. \ref{fig:low_mass_loss}), which is significantly lower than the observed limit. However, a very similar post-main sequence evolution of stars that begin their core helium burning as RSGs regardless of metallicity is not sufficient to explain the metallicity-independent behavior at high overshooting. The two opposing mechanisms that ultimately set the metallicity-independent luminosity cutoffs for RSGs are discussed below. 

\subsection{Metallicity-independent RSG luminosity limit}
\label{sec:HD_limit}
The results discussed above regarding the RSG luminosity limit for the low and high overshooting cases are summarized in Fig. \ref{fig:low_mass_loss}. The solid blue line shows the metallicity independent RSG limit across the three metallicities considered, with an absolute value of $\log (L/L_\odot) \approx 5.5$. For comparison we show the metallicity dependent behavior for the low overshooting case (blue dashed line). We also plot the RSG limits obtained for models that invoke efficient semiconvective mixing to study the time spent in the RSG phase \citep{Higgins2020,Sc2019}.  These limits are adopted from \citet{Higgins2020} (see their Fig. 6) and shown by the red dashed line. 

The figure can be used to qualitatively compare the metallicity trend of the RSG limit while including different types of mixing. The imprint of the metallicity-dependent post-main sequence radius evolution of low overshooting models on the RSG luminosity limit is clearly visible in Fig. \ref{fig:low_mass_loss}. Combined with semiconvective mixing being relevant only at lower metallicities the RSG limit has a Z-dependence, increasing with increasing metallicity.  
Comparing the blue and the red dashed lines, it is evident that the excess mixing in the superadiabatic layers significantly affects the post-main sequence evolution of massive stars. Stars with initial masses beyond the mass range considered in this study ($\gtrsim 40 M_\odot$) are also not expected to evolve as RSGs if the superadiabatic layers are efficiently mixed, keeping the star relatively blue. While we refrain from a complete quantitative comparison between the models due to small  differences in the input physics involved, the general effect of including envelope mixing is drastically lowering the cutoff luminosities (also see Fig. \ref{fig:with_without_MLT++}). 

On the other hand, the higher overshooting mixing results in a very similar pre-RSG evolution of stars for the metallicities considered. These stars begin their core helium burning as RSGs regardless of metallicity wherein physical processes such as mass loss and mixing significantly affect their evolution and set an upper luminosity limit for the cooler supergiants. In contrast, models with lower overshooting show a clear metallicity trend with stars remaining compact at lower metallicities giving a luminosity cutoff that ultimately varies with the metallicity. 

\begin{figure}
    \includegraphics[width = \columnwidth]{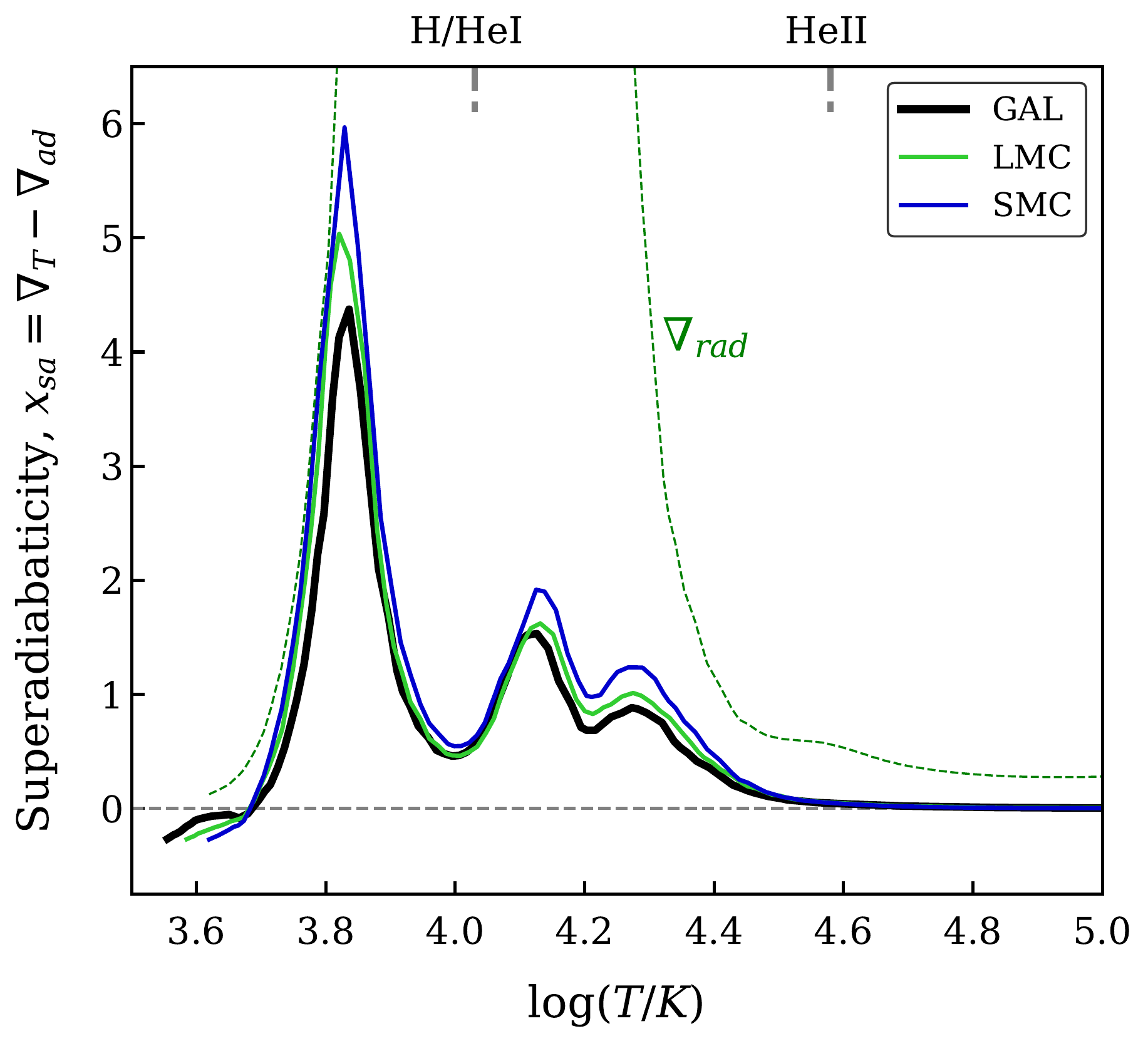}
    \caption{Distribution of superadiabaticity $x_{\text{sa}}$ as a function of temperature inside a $30 M_\odot$ star for three different metallicities: SMC (blue), LMC (green) and Gal (black). The profiles are taken almost halfway through helium burning ($X_c \approx 0.5$) when the stars are evolving as an RSG. The grey dashed line shows exactly zero superadiabaticity and values of $x_{\text{sa}}>0$ are convective regions while $x_{\text{sa}}<0$ are radiative regions. The green dashed line shows the radiative temperature gradient that closely follows the opacity bumps marked at $\log (T_{\text{eff}}/\mathrm{K}) \approx 4$ and $4.6$.} 
    \label{fig:x_sa}
\end{figure}

As previously mentioned, the direct effect of mass loss is to shift the RSG luminosity limit upwards at lower metallicities. Thus, a metallicity independent  observation of  this  limit suggests a process that counteracts the effects of mass loss, pushing it downwards towards lower luminosities at lower metallicities. To identify such a process, we carefully investigate the treatment of superadiabatic mixing by studying the internal temperature gradients as a function of metallicity. For this purpose, we run additional 30 $M_\odot$ models with $f_{\text{ov}} = 0.05$ during the main sequence, but without excess mixing in the superadiabatic layers. In Fig. \ref{fig:x_sa} we show the variation of superadiabaticity $x_{\text{sa}}$ as function of the temperature inside the star, for three different metallicities. Again, these models are not the same as the 30 $M_\odot$ models in Fig. \ref{fig:HRD} that take into account the treatment of excess mixing in the superadiabatic layers. 

The opacities in the bumps located at $\log (T_{\text{eff}}/\mathrm{K}) \approx 5.3$ (iron bump) and $6.2$ (deep iron bump) show a metallicity trend, with smaller peaks at lower metallicities. This is due to the reduction in the iron content whose bound-free transitions are responsible for these bumps. \citet{Sanyal2017} discuss the Z-dependency of envelope inflation with higher metallicity models experiencing inflation effects at lower mass. However, the opacities at cooler bumps, especially the H/He\,\textsc{i} bump at $\log (T_{\text{eff}}/\mathrm{K}) \approx 4$, follow an opposite trend owing to higher hydrogen mass fraction and higher densities causing these opacity peaks to get stronger with decreasing metallicity. In the outermost layers, the temperature gradient required to transport the energy closely follows the radiative temperature gradient as radiative transport gains importance. Just above the hydrogen bump, we have $\nabla_{T} \approx \nabla_{\text{rad}}$ at $\log (T_{\text{eff}}/\mathrm{K}) \approx 3.8$. As $\nabla_{\text{rad}} \sim \chi$, the actual temperature gradient required to transport the radiation increases with decreasing metallicity near the H/He\,\textsc{i} bump. As seen in Fig. \ref{fig:x_sa}, convection becomes more inefficient at lower metallicities as tracked by the variation of the degree of superadiabaticity $x_{\text{sa}}$. These opacity bumps get more pronounced at lower metallicities resulting in more inefficient convection at lower metallicities. Further inside at $\log (T_{\text{eff}}/\mathrm{K}) \approx 4-4.4$, the actual temperature gradient lies between $\nabla_{\text{ad}}$ and $\nabla_{\text{rad}}$ depending on the details of the convection theory used. But as seen from Fig. \ref{fig:x_sa}, the variation of $x_{\text{sa}}$ at these higher temperatures follows a similar metallicity trend. Thus any mixing treatment of inefficient convection that reduces the superadiabaticity would do so as a function of metallicity. In our case, models with SMC metallicity would undergo a larger decrease in $x_{\text{sa}}$ compared to the Galactic models. This results in more mixing in the lower metallicity models keeping them more compact and bluer. This effect of mixing that decreases the RSG limit with decreasing metallicities balances the direct effect of mass loss, thus giving the limit an overall metallicity independent behavior.

\subsection{Effect of mass loss}
\label{section:mass_loss}

Here we discuss the effects of reduced mass loss on our results, by replacing the standard RSG metallicity-independent mass-loss rates of de Jager with the Fe-dependent recipe of \citet{Vink2001} in the yellow supergiant regime ($5500\,\text{K}<T_{\text{eff}}<10\,$kK). The mass loss in the YSG regime are highly uncertain \citep[see][]{Lobel2003, GH2019, Andrews2019, Koumpia2020, Gras2021} and understanding the variation of mass loss as a function of effective temperature is crucial.  The RSG rates by de Jager are still used for cooler temperatures. We compare these two mass-loss recipes for the case of higher overshooting ($f_{\text{ov}} = 0.05$), that can reproduce the luminosity limit. The rates during the main sequence are unchanged, resulting in identical evolution until core hydrogen exhaustion. As seen in Fig \ref{fig:HRD}, the models with higher overshooting at all three metallicities quickly evolve through the Hertzsprung gap phase, beginning their helium burning as an RSG. Thus the redward motion during this quick phase is hardly affected by the lowered mass loss.

This is however not true for the blueward evolution of RSGs, where our models do spend a significant fraction of their helium burning time in the temperature range considered. A reduction in mass loss by almost an order of magnitude in this temperature regime affects the stripping of the envelope. Reduced mass loss results in stars ending their helium burning as yellow supergiants instead of hotter Wolf-Rayet stars, as they are unable to expose their inner layers and evolve bluewards. While the distribution of time spent as an RSG remains \textit{unchanged}, the surface hydrogen mass fraction distributions are different, where models with lowered mass loss retain their envelope and have higher surface hydrogen mass fraction.  We have also tested the effect of having metallicity-dependent mass-loss rates in the RSG regime. We find that by completely removing the de Jager rates, we revert back to a weak dependence on metallicity according to the first order effect of mass loss with the luminosity limit shifting to higher luminosities at lower metallicities (black dashed line in Fig. \ref{fig:low_mass_loss}). This suggests that the metallicity independent nature of the RSG luminosity limit across three metallicities depends both on the treatment of internal mixing as well as mass loss, specifically the temperature of the switch from a metallicity-dependent to a metallicity-independent wind.

\subsection{Overshooting during core helium burning}
\label{sec:he_os}

The effect of overshooting mixing above the helium core after the main sequence on models is often not investigated in stellar evolution studies. Since we primarily focus on mixing processes during the helium burning phase of massive stars, we have also tested the effect of having a higher overshooting mixing above the helium core. Prior to this section, all models implemented an overshooting parameter of $f_{\text{ov}} = 0.01$ above the helium core. We increase this to a value of 0.05 for the study here, which increases the helium core size. Consequently this increases the helium fuel available for burning causing the helium burning time to increase by a small fraction ($\sim 1\%$). In general, this does not affect the fraction of helium burning time spent as RSG, but slightly changes the final mass and the surface abundances at the end of helium burning. Thus, the RSG luminosity limit is not significantly changed by the presence of a larger core during helium burning, and one would expect a very similar distribution as the blue solid line in Fig. \ref{fig:low_mass_loss}. 

\subsection{Constraining the superadiabatic mixing}
\label{section:MLT++EFFECT}

The observed luminosity cutoff at $\log (L/L_\odot) \approx 5.5$ allows to constrain the extra mixing in the superadiabatic regions when radiation pressure in the envelope of these stars begins to dominate. Specifically, we can constrain how close stars should be to radiation pressure dominance for the treatment of superadiabatic mixing to reproduce the observed luminosity cutoff. Accordingly we explore the effects of different sets of MLT++ parameters on the post-main sequence evolution and consequently the RSG limit. We adopt three different sets of parameters as discussed in Appendix \ref{appendix:A} - the "standard" (green) parameter set used to reproduce our results in Sect. \ref{sec:HRD} along with the cases of "weaker"(black) and "stronger"(blue) mixing. The detailed values of the parameters is provided in Appendix \ref{appendix:A} and the color scheme follows Fig. \ref{fig:MLT++PS1}. 

\begin{figure}
    \includegraphics[width = \columnwidth]{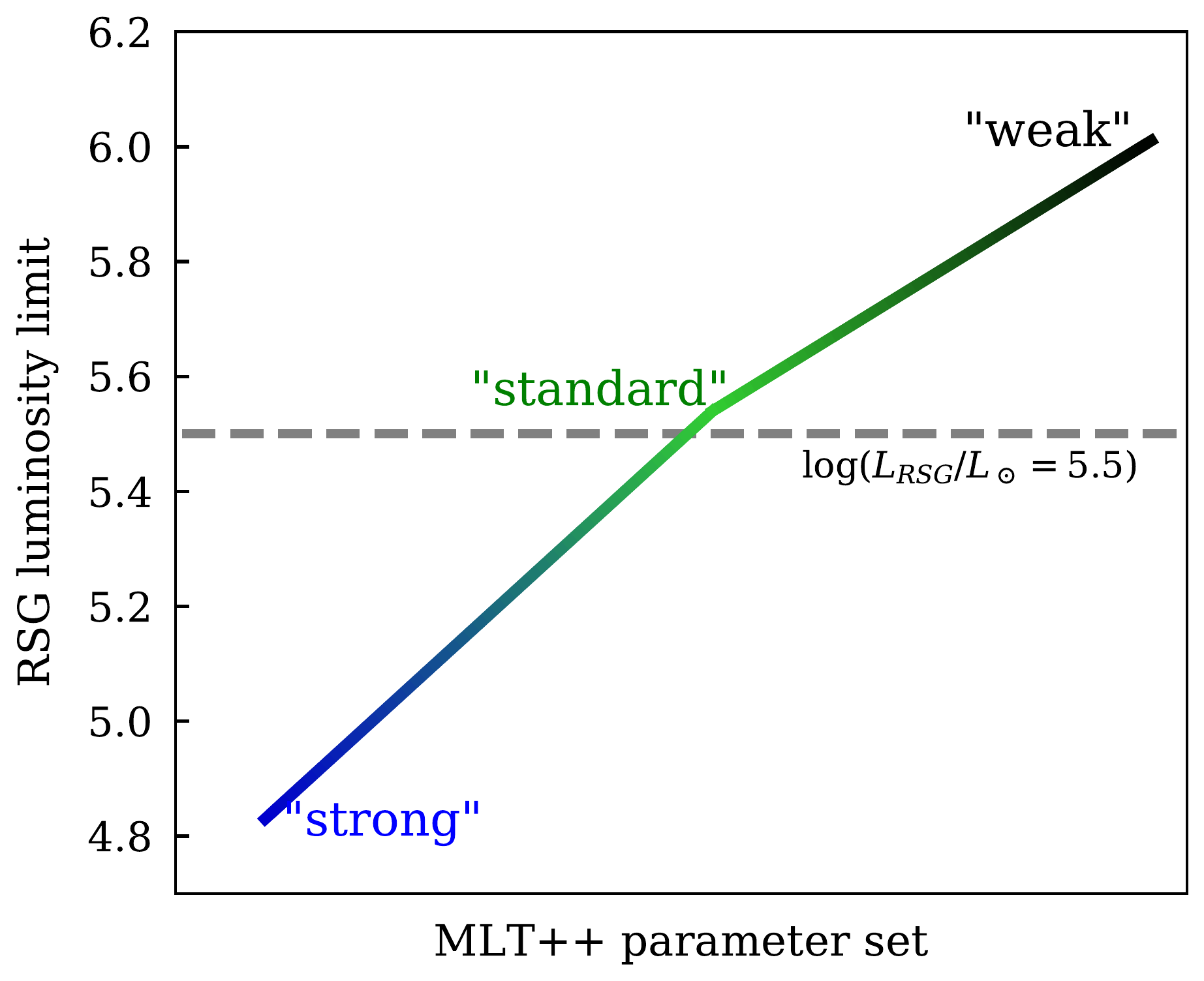}
    \caption{The variation of the RSG luminosity cutoff for the different sets of parameters used. Models at SMC metallicity are used to test the effects of "strong" (blue) and "weak" (black) mixing in the superadiabatic layers where the color scheme follows from Fig. \ref{fig:MLT++PS1}. The grey dashed line represents the observed RSG luminosity limit.}
    \label{fig:smc_varying_parameters}
\end{figure}

The main difference between these models is their closeness to radiation pressure dominance when we choose to efficiently mix the superadiabatic layers instead of allowing them to inflate and develop density inversions. As seen from Fig. \ref{fig:smc_varying_parameters}, the set of parameter values used does indeed affect the location of the cutoff luminosity (solid line), with "stronger" mixing resulting in stars no longer evolving towards the red, and models with "weaker" mixing resulting in RSGs well above the observed constraint.

For the case of "stronger" mixing, the superadiabatic treatment in these core helium burning stars begins when the minimum value of $\beta(r)$ inside the star is as high as $0.7$. This results in blueward evolution of stars with initial masses as low as $10$ $M_\odot$. This significantly lowers the RSG luminosity limit below $\log (L/L_\odot) \approx 5.0$  as seen in Fig. \ref{fig:smc_varying_parameters}. Likewise, the case of "weak" mixing only begins the treatment of superadiabatic mixing when the minimum $\beta$ inside the star is lower than $\approx 0.1$. This causes stars to evolve as RSGs well above the observed constraint of $\log (L/L_\odot) \approx 5.5$. 

The observed luminosity cutoff is reproduced in our models when they are treated for the superadiabatic mixing beginning from minimum $\beta$ values below $\approx 0.2$. While this indicates that stars with initial masses as low as $25 - 30$ $M_\odot$ are affected by this formulation, the full efficiency boost in the superadiabatic treatment ($\alpha = 1$) is not realised until the minimum $\beta$ inside the star falls below $\approx 0.1$. The calculation of $\alpha$ and more details regarding the time smoothing of this efficiency is provided in Appendix \ref{appendix:A}. This test of the parameter space of MLT++ suggests that the "standard" set of parameters chosen in Sect. \ref{section:mixing} not only gives the predicted observed limit, but also sets this limit independent of the metallicity. Thus one can safely disregard the "weaker" and "stronger" part of the excess mixing parameter space, with the former resulting in RSGs with luminosities above the empirical limit, while the latter hardly forming any RSGs. 

\section{Discussion}
\label{sec:disc}

The absolute value of the temperature-independent luminosity cutoff of cooler supergiants and what physical processes set this limit are long standing problems ever since \citet{HD1979} discussed the lack of M supergiants above a certain bolometric magnitude. Metallicity dependent mass loss was initially thought to be the primary factor that sets the limit, where stars with low metallicity are unable to get rid of the massive convective envelope, thus shifting the limit towards higher luminosities. This clearly results in the HD limit being metallicity dependent. Models with lower metallicity are slightly more luminous (opacity effect) at the end of hydrogen exhaustion, and assuming an almost horizontal Hertzsprung gap phase, they produce overall more luminous supergiants compared to higher metallicity models. However, this increase in the luminosity purely as an effect of opacity is insignificant ($\ll 0.1$ dex) compared to the shift due to mass loss. \citet{Davies2018} found, quite surprisingly, lower limits for LMC and SMC metallicities, resulting in a metallicity trend opposite to that expected from the direct effect of mass loss. The lowered limit for the Magellanic Clouds were reproduced in \citet{Higgins2020} using enhanced semiconvective mixing that resulted in the blueward motion of these stars, lowering the limit at these considered metallicities. At higher metallicities, these semiconvective layers disappeared owing to higher mass-loss rates, thus resulting in an indirect effect of mass loss.  

A recent study by \citet{Davies2020} on the luminosities of the most luminous RSGs in the Galaxy suggests a down-revised RSG luminosity limit of $\log (L/L_\odot) \approx 5.5$. If the limit is truly revised down to 5.5 for the Galaxy, then the limit would infact be Z-independent. The metallicity independent behaviour of this limit suggests a physical process that affects stars regardless of their initial metallicity. 

Given that many massive O-type stars are known to be part of a close binary system, one may wonder whether binary Roche-Lobe overflow may be at the heart of this $Z$-independent HD limit. 
However, as approximately 50$\%$ of massive O-type stars were not detected to be in close binaries \citep{Sana2013, Kobul2014}, it seems unlikely that general Hertzsprung-Russell diagram features, such as the HD limit, are solely a binarity effect. It would also raise the question what is so special in binaries at $\log (L/L_\odot) \approx 5.5$ (corresponding to a mass of 30 $M_\odot$) that stars above this mass never turn redwards. 
It is thus more likely that the 
HD feature is at its heart a single star effect, and binarity plays an additional role, in that the expansion and inflation of single stars affects the probability of binary interaction in some unknown fraction of the population. Either way, we need to understand the structure of the envelopes of massive stars.

In this paper we discussed extra mixing in the superadiabatic regions of the star when they are close to their Eddington limit and radiation pressure begins to dominate.  These superadiabatic convective regions are radiation dominated, i.e., the temperature gradient required to transport the energy is given by the radiative temperature gradient. This occurs closer to the surface near the H/HeI opacity bump. This bump becomes stronger as the metallicity decreases, causing the temperature gradient required to transport the energy to increase. Thus the stars at lower metallicities have higher superadiabaticities on average and convective energy transport becomes more inefficient. This would mean stars with different metallicities are subjected to different amounts of mixing in their superadiabatic regions with MLT++, with lower metallicities experiencing higher mixing, helping to keep these stars bluer. This balances the effect of mass loss on the limit by shifting it to lower luminosities at lower metallicities. Although this mixing starts taking effect at roughly the same luminosity across all three metallicities, the boost in efficiency is clearly metallicity dependent. 

Another interesting result shows the effect of mass loss, specifically the highly uncertain YSG mass-loss rates, on the RSG luminosity limit. Mass-loss rates in this regime become important to understand two key points. First, as shown earlier, metallicity independent mass loss in the RSG regime given by the de Jager rates is essential to constrain the behavior of the RSG limit as a function of metallicity. Having a metallicity dependent mass-loss prescription in this regime makes the RSG limit vary with metallicity as expected. This is however a weak function of metallicity ($\lesssim 0.1$ dex in luminosity) as seen from the black dashed line in Fig. \ref{fig:low_mass_loss}. This weak behavior in our models is noticed even with the treatment of the superadiabatic layers, highlighting the importance of dust-driven winds in this regime. In the absence of such a treatment, the limit would further shift upwards at lower metallicity as there would be no balancing effect from the excess mixing, thus giving a strong metallicity trend. Also important to the post-RSG evolution of our models are the YSG mass-loss rates and the location of the switch from metallicity-dependent to independent winds.


Second, the post-RSG evolution of the stars towards the blue by envelope stripping depends on the YSG mass-loss rates. While excess envelope mixing can explain the scarcity of RSGs above a certain luminosity, such a limit on the YSG population still needs to be reconciled with the existing stellar evolution models. As pointed out in a recent work by \citet{Gilkis2021}, current stellar models are unable to explain the lack of cooler supergiants above $\log (L/L_\odot) \approx 5.5$, where "cooler" supergiants collectively represent blue, yellow and red supergiants with effective temperatures less than $15\,$kK. Both mass loss and mixing, in the form of higher convective boundary mixing during main sequence and potentially enhanced mixing during helium burning play an important role in the evolution of these yellow supergiants. The absence of cooler supergiants in this regime hints towards enhanced mixing as well as metallicity independent winds in this regime. While our models can reproduce the metallicity-independent behaviour of the HD limit for the red supergiants, the so-called yellow supergiant problem still remains unresolved.

An interesting consequence of having higher convective boundary mixing above the core during the main sequence is the increase in the main sequence width. \citet{Vink2010} discussed the possibility of "bi-stability braking" (BSB) as a possible explanation for the drop in the rotation rates of massive stars below $T_{\text{eff}} =  22000\,$K. The BSB mechanism is applicable for stars that expand to cooler temperatures below $T_{\text{eff}} =  22000\,$K during main sequence. A wider main sequence is favored at higher overshooting mixing as seen in Fig. \ref{fig:HRD}. The applicability of the BSB  mechanism is thus pushed to lower masses at higher overshooting. This would also have consequences on the evolutionary status of B supergiants. The increased main sequence width could potentially explain the large number of B supergiants, as they are still main sequence objects.

However for high overshooting ($\alpha_{\text{ov}} \simeq 0.3-0.5$; \citet{Vink2010, Castro2014}), after core hydrogen exhaustion the stars would quickly traverse the Hertzsprung gap phase and effectively start core helium burning as RSGs (for masses below $\sim 25 \,M_\odot$). This behavior being almost metallicity independent is crucial for reproducing the metallicity independent RSG limit. This however suggests that with higher overshooting our models predict almost no pre-RSG, cooler blue and yellow supergiants, which is not the case for the lower overshooting models as seen from Fig. \ref{fig:HRD}. One possible resolution for this problem is to have a population of massive stars with a wide range of overshooting above the core, which is already hinted for the case of lower mass stars, upto $\sim 25 \; \text{M}_\odot$ with $\alpha_{\text{ov}} = 0 - 0.44$ from asteroseismological studies \citep{Bowman2020}. The RSG luminosity limit is then set by the higher overshooting models, while still producing pre-RSG cooler BSGs and YSGs. Of course, a quantitative analysis of populations of these supergiants, such as the B/R ratio as a function of luminosity, ultimately should come from detailed population synthesis models.

The  current observational constraints on the luminosity of the coolest supergiants for different metallicities offers an opportunity to inform us the effects of different physical processes as a function of metallicity. As discussed in Sect. \ref{sec:HD_limit}, any mixing treatment of inefficient convection that reduces the superadiabaticity would do so as a function of metallicity, inducing more mixing at lower  metallicity. This is the effect of excess envelope mixing that counteracts the direct effect of mass loss. If the limit is truly metallicity independent beyond the range $\frac{1}{5} Z_\odot \lesssim  Z  \lesssim  Z_\odot$ considered in this study, this would mean the effects of excess envelope mixing and mass loss are indeed balancing each other out. However, if new observational constraints are available that strongly favours a metallicity dependent limit, this would suggest one of the aforementioned processes to dominate over the other. A shift in the limits to higher luminosities at sub-solar metallicities would favour the effects of mass loss over excess mixing. Likewise, if the RSG limit at solar and super-solar metallicities with future observations are indeed at $\log (L/L_\odot) \approx 5.8$, this would suggest that the effects of mixing would dominate over mass loss.

\section{Conclusions}
\label{sec:conc}

In this study, we have produced a grid of stellar evolution models at three initial metallicities and two different values of overshooting mixing above the hydrogen core to investigate the effect of excess mixing in the superadiabatic layers when stars evolve close to their Eddington limit on the RSG upper luminosity limit. We extensively probe non-rotating models, however the argument regarding a luminosity cutoff for the cooler supergiants extends to rotating models as well with the slow rotators shifting upwards and bluewards in the HRD, while the fast rotators undergo chemical homogeneous evolution and do not influence the cutoff luminosity. 

Our higher overshooting models are able to reproduce the observed metallicity-independent RSG luminosity limit at $\log (L/L_\odot) \approx 5.5$ for the metallicities considered in this study. Metallicity has minimal effect on the post-main sequence evolution of models with higher overshooting mixing wherein they begin core helium burning as RSGs regardless of metallicity. The metallicity-independent RSG limit is ultimately set by the conditions of mass loss and mixing in the RSG regime. While mass loss tends to shift this limit to higher luminosities at lower metallicities, we find that the excess mixing in the superadiabatic layers of stars has an opposing effect, thus balancing the direct effect of mass loss, giving a metallicity-independent limit. The metallicity-independent dust-driven winds  are important in the RSG regime, in the absence of which the RSG limit becomes a weak function of metallicity and increases with decreasing metallicity. Both mass-loss rates and mixing processes and their efficiencies in the RSG regime are highly uncertain and are important to understand how they influence the luminosity limit of the coolest supergiants as a function of metallicity. 

 While at high values of overshooting we can reproduce the metallicity independent RSG limit at the observed luminosity, these models predict no pre-RSG yellow supergiants between luminosities $\log (L/L_\odot) \approx 4.5 - 5$. This can be at least qualitatively explained with a population of stars with a range of overshooting, with the higher overshooting models setting the RSG limit and the lower overshooting models responsible for the formation of blue supergiant population. The high overshooting during main sequence also has consequences on the mass range of applicability of BSB and could potentially explain the large number of B supergiants.

The yellow supergiant problem \citep[see also][]{Gilkis2021} still remains unresolved wherein the current stellar evolution models over-predict the number of cooler supergiants above $\log (L/L_\odot) \approx 5.5$.
Higher mass-loss rates and efficient mixing in this regime could potentially solve this problem. Careful consideration of the wind physics and mixing along with observational constraints on mass-loss rates in this regime can help tackle this problem. 

Finally, if future observational constraints either confirm a Z-independent limit or strongly favours a limit that varies with metallicity, it informs us about the physical processes inside the star and which effects dominate over the other.

\section*{Acknowledgements}

We thank the anonymous referee for constructive comments that helped improve the paper. We also thank the MESA developers for making their stellar evolution code publicly available.
JSV and ERH are supported by STFC funding under grant number ST/R000565/1.
AACS is an \"{O}pik Research Fellow.


\section*{Data Availability}

The data underlying this article will be shared on reasonable request
to the corresponding author.




\bibliographystyle{mnras}
\bibliography{References} 

\begin{thebibliography}{}
\makeatletter
\relax
\def\mn@urlcharsother{\let\do\@makeother \do\$\do\&\do\#\do\^\do\_\do\%\do\~}
\def\mn@doi{\begingroup\mn@urlcharsother \@ifnextchar [ {\mn@doi@}
  {\mn@doi@[]}}
\def\mn@doi@[#1]#2{\def\@tempa{#1}\ifx\@tempa\@empty \href
  {http://dx.doi.org/#2} {doi:#2}\else \href {http://dx.doi.org/#2} {#1}\fi
  \endgroup}
\def\mn@eprint#1#2{\mn@eprint@#1:#2::\@nil}
\def\mn@eprint@arXiv#1{\href {http://arxiv.org/abs/#1} {{\tt arXiv:#1}}}
\def\mn@eprint@dblp#1{\href {http://dblp.uni-trier.de/rec/bibtex/#1.xml}
  {dblp:#1}}
\def\mn@eprint@#1:#2:#3:#4\@nil{\def\@tempa {#1}\def\@tempb {#2}\def\@tempc
  {#3}\ifx \@tempc \@empty \let \@tempc \@tempb \let \@tempb \@tempa \fi \ifx
  \@tempb \@empty \def\@tempb {arXiv}\fi \@ifundefined
  {mn@eprint@\@tempb}{\@tempb:\@tempc}{\expandafter \expandafter \csname
  mn@eprint@\@tempb\endcsname \expandafter{\@tempc}}}

\bibitem[\protect\citeauthoryear{{Abbott}}{{Abbott}}{1982}]{Abbott1982}
{Abbott} D.~C.,  1982, \mn@doi [\apj] {10.1086/160166}, \href
  {https://ui.adsabs.harvard.edu/abs/1982ApJ...259..282A} {259, 282}

\bibitem[\protect\citeauthoryear{{Andrews}, {Fenech}, {Prinja}, {Clark}  \&
  {Hindson}}{{Andrews} et~al.}{2019}]{Andrews2019}
{Andrews} H.,  {Fenech} D.,  {Prinja} R.~K.,  {Clark} J.~S.,   {Hindson} L.,
  2019, \mn@doi [\aap] {10.1051/0004-6361/201936256}, \href
  {https://ui.adsabs.harvard.edu/abs/2019A&A...632A..38A} {632, A38}

\bibitem[\protect\citeauthoryear{{Bestenlehner} et~al.,}{{Bestenlehner}
  et~al.}{2014}]{Best2014}
{Bestenlehner} J.~M.,  et~al., 2014, \mn@doi [\aap]
  {10.1051/0004-6361/201423643}, \href
  {https://ui.adsabs.harvard.edu/abs/2014A&A...570A..38B} {570, A38}

\bibitem[\protect\citeauthoryear{{Bonaca} et~al.,}{{Bonaca}
  et~al.}{2012}]{Bonaca2012}
{Bonaca} A.,  et~al., 2012, \mn@doi [\apjl] {10.1088/2041-8205/755/1/L12},
  \href {https://ui.adsabs.harvard.edu/abs/2012ApJ...755L..12B} {755, L12}

\bibitem[\protect\citeauthoryear{{Bowman}}{{Bowman}}{2020}]{Bowman2020}
{Bowman} D.~M.,  2020, \mn@doi [Frontiers in Astronomy and Space Sciences]
  {10.3389/fspas.2020.578584}, \href
  {https://ui.adsabs.harvard.edu/abs/2020FrASS...7...70B} {7, 70}

\bibitem[\protect\citeauthoryear{{Brott} et~al.,}{{Brott}
  et~al.}{2011}]{Brott2011}
{Brott} I.,  et~al., 2011, \mn@doi [\aap] {10.1051/0004-6361/201016113}, \href
  {https://ui.adsabs.harvard.edu/abs/2011A&A...530A.115B} {530, A115}

\bibitem[\protect\citeauthoryear{{Castro}, {Fossati}, {Langer},
  {Sim{\'o}n-D{\'\i}az}, {Schneider}  \& {Izzard}}{{Castro}
  et~al.}{2014}]{Castro2014}
{Castro} N.,  {Fossati} L.,  {Langer} N.,  {Sim{\'o}n-D{\'\i}az} S.,
  {Schneider} F.~R.~N.,   {Izzard} R.~G.,  2014, \mn@doi [\aap]
  {10.1051/0004-6361/201425028}, \href
  {https://ui.adsabs.harvard.edu/abs/2014A&A...570L..13C} {570, L13}

\bibitem[\protect\citeauthoryear{{Choi}, {Dotter}, {Conroy}, {Cantiello},
  {Paxton}  \& {Johnson}}{{Choi} et~al.}{2016}]{CHOI16}
{Choi} J.,  {Dotter} A.,  {Conroy} C.,  {Cantiello} M.,  {Paxton} B.,
  {Johnson} B.~D.,  2016, \mn@doi [\apj] {10.3847/0004-637X/823/2/102}, \href
  {https://ui.adsabs.harvard.edu/abs/2016ApJ...823..102C} {823, 102}

\bibitem[\protect\citeauthoryear{{Cox} \& {Giuli}}{{Cox} \&
  {Giuli}}{1968}]{MLT68}
{Cox} J.~P.,  {Giuli} R.~T.,  1968, {Principles of stellar structure}

\bibitem[\protect\citeauthoryear{{Davies} \& {Beasor}}{{Davies} \&
  {Beasor}}{2020}]{Davies2020}
{Davies} B.,  {Beasor} E.~R.,  2020, \mn@doi [\mnras] {10.1093/mnras/staa174},
  \href {https://ui.adsabs.harvard.edu/abs/2020MNRAS.493..468D} {493, 468}

\bibitem[\protect\citeauthoryear{{Davies}, {Crowther}  \& {Beasor}}{{Davies}
  et~al.}{2018}]{Davies2018}
{Davies} B.,  {Crowther} P.~A.,   {Beasor} E.~R.,  2018, \mn@doi [\mnras]
  {10.1093/mnras/sty1302}, \href
  {https://ui.adsabs.harvard.edu/abs/2018MNRAS.478.3138D} {478, 3138}

\bibitem[\protect\citeauthoryear{{Drout}, {Massey}, {Meynet}, {Tokarz}  \&
  {Caldwell}}{{Drout} et~al.}{2009}]{Drout2009}
{Drout} M.~R.,  {Massey} P.,  {Meynet} G.,  {Tokarz} S.,   {Caldwell} N.,
  2009, \mn@doi [\apj] {10.1088/0004-637X/703/1/441}, \href
  {https://ui.adsabs.harvard.edu/abs/2009ApJ...703..441D} {703, 441}

\bibitem[\protect\citeauthoryear{{Gilkis}, {Shenar}, {Ramachandran}, {Jermyn},
  {Mahy}, {Oskinova}, {Arcavi}  \& {Sana}}{{Gilkis} et~al.}{2021}]{Gilkis2021}
{Gilkis} A.,  {Shenar} T.,  {Ramachandran} V.,  {Jermyn} A.~S.,  {Mahy} L.,
  {Oskinova} L.~M.,  {Arcavi} I.,   {Sana} H.,  2021, \mn@doi [\mnras]
  {10.1093/mnras/stab383}, \href
  {https://ui.adsabs.harvard.edu/abs/2021MNRAS.tmp..422G} {}

\bibitem[\protect\citeauthoryear{{Gordon} \& {Humphreys}}{{Gordon} \&
  {Humphreys}}{2019}]{GH2019}
{Gordon} M.~S.,  {Humphreys} R.~M.,  2019, \mn@doi [Galaxies]
  {10.3390/galaxies7040092}, \href
  {https://ui.adsabs.harvard.edu/abs/2019Galax...7...92G} {7, 92}

\bibitem[\protect\citeauthoryear{{Gr{\"a}fener}, {Owocki}  \&
  {Vink}}{{Gr{\"a}fener} et~al.}{2012}]{Graf2012}
{Gr{\"a}fener} G.,  {Owocki} S.~P.,   {Vink} J.~S.,  2012, \mn@doi [\aap]
  {10.1051/0004-6361/201117497}, \href
  {https://ui.adsabs.harvard.edu/abs/2012A&A...538A..40G} {538, A40}

\bibitem[\protect\citeauthoryear{{Grassitelli}, {Langer}, {Mackey},
  {Gr{\"a}fener}, {Grin}, {Sander}  \& {Vink}}{{Grassitelli}
  et~al.}{2021}]{Gras2021}
{Grassitelli} L.,  {Langer} N.,  {Mackey} J.,  {Gr{\"a}fener} G.,  {Grin}
  N.~J.,  {Sander} A.~A.~C.,   {Vink} J.~S.,  2021, \mn@doi [\aap]
  {10.1051/0004-6361/202038298}, \href
  {https://ui.adsabs.harvard.edu/abs/2021A&A...647A..99G} {647, A99}

\bibitem[\protect\citeauthoryear{{Grevesse} \& {Sauval}}{{Grevesse} \&
  {Sauval}}{1998}]{GS98}
{Grevesse} N.,  {Sauval} A.~J.,  1998, \mn@doi [\ssr]
  {10.1023/A:1005161325181}, \href
  {https://ui.adsabs.harvard.edu/abs/1998SSRv...85..161G} {85, 161}

\bibitem[\protect\citeauthoryear{{Heger}, {Langer}  \& {Woosley}}{{Heger}
  et~al.}{2000}]{Heger2000}
{Heger} A.,  {Langer} N.,   {Woosley} S.~E.,  2000, \mn@doi [\apj]
  {10.1086/308158}, \href
  {https://ui.adsabs.harvard.edu/abs/2000ApJ...528..368H} {528, 368}

\bibitem[\protect\citeauthoryear{{Herwig}}{{Herwig}}{2000}]{Herwig2000}
{Herwig} F.,  2000, \aap, \href
  {https://ui.adsabs.harvard.edu/abs/2000A&A...360..952H} {360, 952}

\bibitem[\protect\citeauthoryear{{Higgins} \& {Vink}}{{Higgins} \&
  {Vink}}{2019}]{Higgins2019}
{Higgins} E.~R.,  {Vink} J.~S.,  2019, \mn@doi [\aap]
  {10.1051/0004-6361/201834123}, \href
  {https://ui.adsabs.harvard.edu/abs/2019A&A...622A..50H} {622, A50}

\bibitem[\protect\citeauthoryear{Higgins \& Vink}{Higgins \&
  Vink}{2020}]{Higgins2020}
Higgins E.,  Vink J.~S.,  2020, Astronomy and Astrophysics, 635, 1937374

\bibitem[\protect\citeauthoryear{{Humphreys} \& {Davidson}}{{Humphreys} \&
  {Davidson}}{1979}]{HD1979}
{Humphreys} R.~M.,  {Davidson} K.,  1979, \mn@doi [\apj] {10.1086/157301},
  \href {https://ui.adsabs.harvard.edu/abs/1979ApJ...232..409H} {232, 409}

\bibitem[\protect\citeauthoryear{{Ishii}, {Ueno}  \& {Kato}}{{Ishii}
  et~al.}{1999}]{Ishii1999}
{Ishii} M.,  {Ueno} M.,   {Kato} M.,  1999, \mn@doi [\pasj]
  {10.1093/pasj/51.4.417}, \href
  {https://ui.adsabs.harvard.edu/abs/1999PASJ...51..417I} {51, 417}

\bibitem[\protect\citeauthoryear{{Jiang}, {Cantiello}, {Bildsten}, {Quataert}
  \& {Blaes}}{{Jiang} et~al.}{2015}]{Jiang2015}
{Jiang} Y.-F.,  {Cantiello} M.,  {Bildsten} L.,  {Quataert} E.,   {Blaes} O.,
  2015, \mn@doi [\apj] {10.1088/0004-637X/813/1/74}, \href
  {https://ui.adsabs.harvard.edu/abs/2015ApJ...813...74J} {813, 74}

\bibitem[\protect\citeauthoryear{{Klencki}, {Nelemans}, {Istrate}  \&
  {Pols}}{{Klencki} et~al.}{2020}]{Klencki2020}
{Klencki} J.,  {Nelemans} G.,  {Istrate} A.~G.,   {Pols} O.,  2020, \mn@doi
  [\aap] {10.1051/0004-6361/202037694}, \href
  {https://ui.adsabs.harvard.edu/abs/2020A&A...638A..55K} {638, A55}

\bibitem[\protect\citeauthoryear{{Klencki}, {Nelemans}, {Istrate}  \&
  {Chruslinska}}{{Klencki} et~al.}{2021}]{Klencki2021}
{Klencki} J.,  {Nelemans} G.,  {Istrate} A.~G.,   {Chruslinska} M.,  2021,
  \mn@doi [\aap] {10.1051/0004-6361/202038707}, \href
  {https://ui.adsabs.harvard.edu/abs/2021A&A...645A..54K} {645, A54}

\bibitem[\protect\citeauthoryear{{Kobulnicky} et~al.,}{{Kobulnicky}
  et~al.}{2014}]{Kobul2014}
{Kobulnicky} H.~A.,  et~al., 2014, \mn@doi [\apjs]
  {10.1088/0067-0049/213/2/34}, \href
  {https://ui.adsabs.harvard.edu/abs/2014ApJS..213...34K} {213, 34}

\bibitem[\protect\citeauthoryear{{K{\"o}hler} et~al.,}{{K{\"o}hler}
  et~al.}{2015}]{Koehler2015}
{K{\"o}hler} K.,  et~al., 2015, \mn@doi [\aap] {10.1051/0004-6361/201424356},
  \href {https://ui.adsabs.harvard.edu/abs/2015A&A...573A..71K} {573, A71}

\bibitem[\protect\citeauthoryear{{Koumpia} et~al.,}{{Koumpia}
  et~al.}{2020}]{Koumpia2020}
{Koumpia} E.,  et~al., 2020, \mn@doi [\aap] {10.1051/0004-6361/201936177},
  \href {https://ui.adsabs.harvard.edu/abs/2020A&A...635A.183K} {635, A183}

\bibitem[\protect\citeauthoryear{{Lamers} \& {Fitzpatrick}}{{Lamers} \&
  {Fitzpatrick}}{1988}]{Lamers1988}
{Lamers} H. J.~G.~L.~M.,  {Fitzpatrick} E.~L.,  1988, \mn@doi [\apj]
  {10.1086/165894}, \href
  {https://ui.adsabs.harvard.edu/abs/1988ApJ...324..279L} {324, 279}

\bibitem[\protect\citeauthoryear{{Lobel} et~al.,}{{Lobel}
  et~al.}{2003}]{Lobel2003}
{Lobel} A.,  et~al., 2003, \mn@doi [\apj] {10.1086/345503}, \href
  {https://ui.adsabs.harvard.edu/abs/2003ApJ...583..923L} {583, 923}

\bibitem[\protect\citeauthoryear{{Maeder}}{{Maeder}}{1987}]{Maeder1987}
{Maeder} A.,  1987, \aap, \href
  {https://ui.adsabs.harvard.edu/abs/1987A&A...178..159M} {178, 159}

\bibitem[\protect\citeauthoryear{{M{\"u}ller} \& {Vink}}{{M{\"u}ller} \&
  {Vink}}{2014}]{Muller2014}
{M{\"u}ller} P.~E.,  {Vink} J.~S.,  2014, \mn@doi [\aap]
  {10.1051/0004-6361/201323031}, \href
  {https://ui.adsabs.harvard.edu/abs/2014A&A...564A..57M} {564, A57}

\bibitem[\protect\citeauthoryear{{Nugis} \& {Lamers}}{{Nugis} \&
  {Lamers}}{2000}]{NL2000}
{Nugis} T.,  {Lamers} H.~J.~G.~L.~M.,  2000, \aap, \href
  {https://ui.adsabs.harvard.edu/abs/2000A&A...360..227N} {360, 227}

\bibitem[\protect\citeauthoryear{{Owocki}}{{Owocki}}{2015}]{Owocki2015}
{Owocki} S.~P.,  2015, {Instabilities in the Envelopes and Winds of Very
  Massive Stars}.
p.~113, \mn@doi{10.1007/978-3-319-09596-7_5}

\bibitem[\protect\citeauthoryear{{Paxton}, {Bildsten}, {Dotter}, {Herwig},
  {Lesaffre}  \& {Timmes}}{{Paxton} et~al.}{2011}]{MESA11}
{Paxton} B.,  {Bildsten} L.,  {Dotter} A.,  {Herwig} F.,  {Lesaffre} P.,
  {Timmes} F.,  2011, \mn@doi [\apjs] {10.1088/0067-0049/192/1/3}, \href
  {https://ui.adsabs.harvard.edu/abs/2011ApJS..192....3P} {192, 3}

\bibitem[\protect\citeauthoryear{{Paxton} et~al.,}{{Paxton}
  et~al.}{2013}]{MESA13}
{Paxton} B.,  et~al., 2013, \mn@doi [\apjs] {10.1088/0067-0049/208/1/4}, \href
  {https://ui.adsabs.harvard.edu/abs/2013ApJS..208....4P} {208, 4}

\bibitem[\protect\citeauthoryear{{Paxton} et~al.,}{{Paxton}
  et~al.}{2015}]{MESA15}
{Paxton} B.,  et~al., 2015, \mn@doi [\apjs] {10.1088/0067-0049/220/1/15}, \href
  {https://ui.adsabs.harvard.edu/abs/2015ApJS..220...15P} {220, 15}

\bibitem[\protect\citeauthoryear{{Paxton} et~al.,}{{Paxton}
  et~al.}{2018}]{MESA18}
{Paxton} B.,  et~al., 2018, \mn@doi [\apjs] {10.3847/1538-4365/aaa5a8}, \href
  {https://ui.adsabs.harvard.edu/abs/2018ApJS..234...34P} {234, 34}

\bibitem[\protect\citeauthoryear{{Paxton} et~al.,}{{Paxton}
  et~al.}{2019}]{MESA19}
{Paxton} B.,  et~al., 2019, \mn@doi [\apjs] {10.3847/1538-4365/ab2241}, \href
  {https://ui.adsabs.harvard.edu/abs/2019ApJS..243...10P} {243, 10}

\bibitem[\protect\citeauthoryear{{Sana} et~al.,}{{Sana}
  et~al.}{2013}]{Sana2013}
{Sana} H.,  et~al., 2013, \mn@doi [\aap] {10.1051/0004-6361/201219621}, \href
  {https://ui.adsabs.harvard.edu/abs/2013A&A...550A.107S} {550, A107}

\bibitem[\protect\citeauthoryear{{Sanyal}, {Grassitelli}, {Langer}  \&
  {Bestenlehner}}{{Sanyal} et~al.}{2015}]{Sanyal2015}
{Sanyal} D.,  {Grassitelli} L.,  {Langer} N.,   {Bestenlehner} J.~M.,  2015,
  \mn@doi [\aap] {10.1051/0004-6361/201525945}, \href
  {https://ui.adsabs.harvard.edu/abs/2015A&A...580A..20S} {580, A20}

\bibitem[\protect\citeauthoryear{{Sanyal}, {Langer}, {Sz{\'e}csi}, {-C Yoon}
  \& {Grassitelli}}{{Sanyal} et~al.}{2017}]{Sanyal2017}
{Sanyal} D.,  {Langer} N.,  {Sz{\'e}csi} D.,  {-C Yoon} S.,   {Grassitelli} L.,
   2017, \mn@doi [\aap] {10.1051/0004-6361/201629612}, \href
  {https://ui.adsabs.harvard.edu/abs/2017A&A...597A..71S} {597, A71}

\bibitem[\protect\citeauthoryear{{Schootemeijer}, {Langer}, {Grin}  \&
  {Wang}}{{Schootemeijer} et~al.}{2019a}]{Abel2019}
{Schootemeijer} A.,  {Langer} N.,  {Grin} N.~J.,   {Wang} C.,  2019a, \mn@doi
  [\aap] {10.1051/0004-6361/201935046}, \href
  {https://ui.adsabs.harvard.edu/abs/2019A&A...625A.132S} {625, A132}

\bibitem[\protect\citeauthoryear{{Schootemeijer}, {Langer}, {Grin}  \&
  {Wang}}{{Schootemeijer} et~al.}{2019b}]{Sc2019}
{Schootemeijer} A.,  {Langer} N.,  {Grin} N.~J.,   {Wang} C.,  2019b, \mn@doi
  [\aap] {10.1051/0004-6361/201935046}, \href
  {https://ui.adsabs.harvard.edu/abs/2019A&A...625A.132S} {625, A132}

\bibitem[\protect\citeauthoryear{{Schwarzschild} \& {H{\"a}rm}}{{Schwarzschild}
  \& {H{\"a}rm}}{1958}]{SH1958}
{Schwarzschild} M.,  {H{\"a}rm} R.,  1958, \mn@doi [\apj] {10.1086/146548},
  \href {https://ui.adsabs.harvard.edu/abs/1958ApJ...128..348S} {128, 348}

\bibitem[\protect\citeauthoryear{{Shaviv}}{{Shaviv}}{2000}]{Shaviv2000}
{Shaviv} N.~J.,  2000, \mn@doi [\apjl] {10.1086/312585}, \href
  {https://ui.adsabs.harvard.edu/abs/2000ApJ...532L.137S} {532, L137}

\bibitem[\protect\citeauthoryear{Song, Alexeeva  \& Zhao}{Song
  et~al.}{2020}]{Song2020}
Song N.,  Alexeeva S.,   Zhao G.,  2020, \mn@doi [Research in Astronomy and
  Astrophysics] {10.1088/1674-4527/20/8/121}, 20, 121

\bibitem[\protect\citeauthoryear{{Viani}, {Basu}, {Ong J.}, {Bonaca}  \&
  {Chaplin}}{{Viani} et~al.}{2018}]{Viani2018}
{Viani} L.~S.,  {Basu} S.,  {Ong J.} M.~J.,  {Bonaca} A.,   {Chaplin} W.~J.,
  2018, \mn@doi [\apj] {10.3847/1538-4357/aab7eb}, \href
  {https://ui.adsabs.harvard.edu/abs/2018ApJ...858...28V} {858, 28}

\bibitem[\protect\citeauthoryear{{Vink}, {de Koter}  \& {Lamers}}{{Vink}
  et~al.}{1999}]{Vink1999}
{Vink} J.~S.,  {de Koter} A.,   {Lamers} H.~J.~G.~L.~M.,  1999, \aap, \href
  {https://ui.adsabs.harvard.edu/abs/1999A&A...350..181V} {350, 181}

\bibitem[\protect\citeauthoryear{{Vink}, {de Koter}  \& {Lamers}}{{Vink}
  et~al.}{2001}]{Vink2001}
{Vink} J.~S.,  {de Koter} A.,   {Lamers} H.~J.~G.~L.~M.,  2001, \mn@doi [\aap]
  {10.1051/0004-6361:20010127}, \href
  {https://ui.adsabs.harvard.edu/abs/2001A&A...369..574V} {369, 574}

\bibitem[\protect\citeauthoryear{{Vink}, {Brott}, {Gr{\"a}fener}, {Langer}, {de
  Koter}  \& {Lennon}}{{Vink} et~al.}{2010}]{Vink2010}
{Vink} J.~S.,  {Brott} I.,  {Gr{\"a}fener} G.,  {Langer} N.,  {de Koter} A.,
  {Lennon} D.~J.,  2010, \mn@doi [\aap] {10.1051/0004-6361/201014205}, \href
  {https://ui.adsabs.harvard.edu/abs/2010A&A...512L...7V} {512, L7}

\bibitem[\protect\citeauthoryear{{Vink}, {Muijres}, {Anthonisse}, {de Koter},
  {Gr{\"a}fener}  \& {Langer}}{{Vink} et~al.}{2011}]{Vink2011}
{Vink} J.~S.,  {Muijres} L.~E.,  {Anthonisse} B.,  {de Koter} A.,
  {Gr{\"a}fener} G.,   {Langer} N.,  2011, \mn@doi [\aap]
  {10.1051/0004-6361/201116614}, \href
  {https://ui.adsabs.harvard.edu/abs/2011A&A...531A.132V} {531, A132}

\bibitem[\protect\citeauthoryear{{Vink}, {Higgins}, {Sander}  \&
  {Sabhahit}}{{Vink} et~al.}{2020}]{Vink2020}
{Vink} J.~S.,  {Higgins} E.~R.,  {Sander} A. A.~C.,   {Sabhahit} G.~N.,  2020,
  arXiv e-prints, \href {https://ui.adsabs.harvard.edu/abs/2020arXiv201011730V}
  {p. arXiv:2010.11730}

\bibitem[\protect\citeauthoryear{{Woosley} \& {Heger}}{{Woosley} \&
  {Heger}}{2006}]{WH2006}
{Woosley} S.~E.,  {Heger} A.,  2006, \mn@doi [\apj] {10.1086/498500}, \href
  {https://ui.adsabs.harvard.edu/abs/2006ApJ...637..914W} {637, 914}

\bibitem[\protect\citeauthoryear{{Yoon} \& {Langer}}{{Yoon} \&
  {Langer}}{2005}]{YL2005}
{Yoon} S.~C.,  {Langer} N.,  2005, \mn@doi [\aap] {10.1051/0004-6361:20054030},
  \href {https://ui.adsabs.harvard.edu/abs/2005A&A...443..643Y} {443, 643}

\bibitem[\protect\citeauthoryear{{Yusof} et~al.,}{{Yusof}
  et~al.}{2013}]{Yusof2013}
{Yusof} N.,  et~al., 2013, \mn@doi [\mnras] {10.1093/mnras/stt794}, \href
  {https://ui.adsabs.harvard.edu/abs/2013MNRAS.433.1114Y} {433, 1114}

\bibitem[\protect\citeauthoryear{{de Jager}, {Nieuwenhuijzen}  \& {van der
  Hucht}}{{de Jager} et~al.}{1988}]{deJager1988}
{de Jager} C.,  {Nieuwenhuijzen} H.,   {van der Hucht} K.~A.,  1988, \aaps,
  \href {https://ui.adsabs.harvard.edu/abs/1988A&AS...72..259D} {72, 259}

\makeatother
\end{thebibliography}




\appendix

\section{Overshooting prescriptions in MESA}
\label{appendix:C}

\begin{figure}
    \includegraphics[width = \columnwidth]{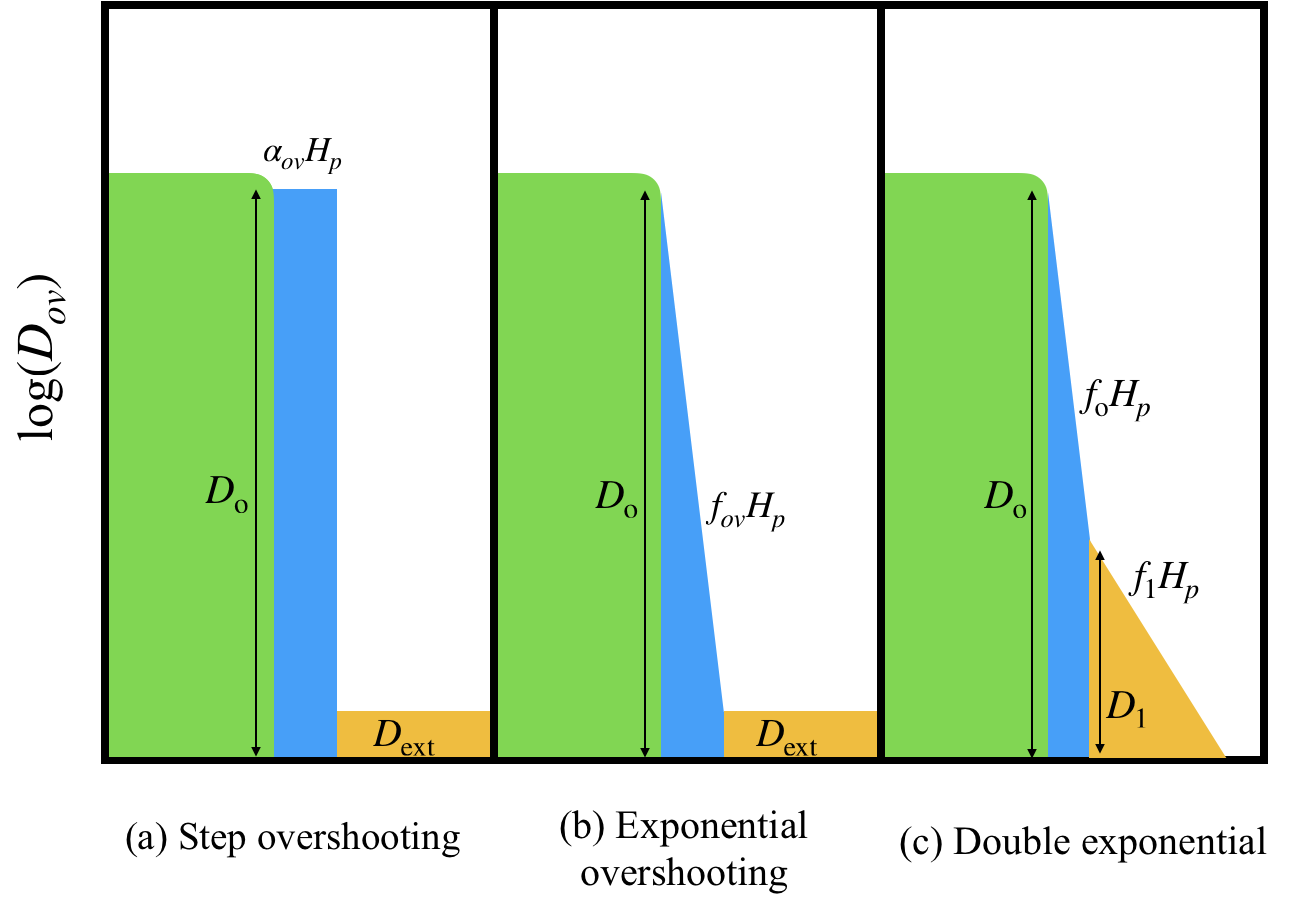}
    \caption{Different overshooting prescriptions available in MESA. The convective regions are shown in green and the overshooting regions in blue. In case of the step and exponential profiles, the radiative regions with a constant diffusion coefficient $D_{\text{ext}}$ is shown in orange. In case of the double exponential case, the radiative diffusion coefficient is replaced by the second exponential profile (also shown in orange). } 
    \label{fig:os_prescription}
\end{figure}

The different overshooting prescriptions available in MESA are discussed below. We also compare the parameters of step and exponential prescriptions that result in similar main sequence evolution.
MESA uses the following diffusion profiles above the convective core to describe the variation of the diffusion coefficient $D_{\text{ov}}$. In all these cases, the overshooting region begins just below the convective boundary from $r_\text{o} = r_\text{b} - \delta r$, where $r_\text{b}$ is the radius where $\nabla_{\text{rad}} = \nabla_{\text{ad}}$. The exact convective boundary is not taken as the starting point of the convective regions because the diffusion coefficient approaches zero at $r_\text{b}$. All prescriptions here are mentioned for overshooting above the convective core and can be extended for convective shells as well. 
\begin{enumerate}
 \item \textbf{Step overshooting}: The simplest of the prescription where the overshooting is described by a constant diffusion coefficient from $r_\text{o}$  out to a fixed distance effectively increasing the size of the core:
\begin{equation}
\begin{array}{c@{\qquad}c}
D_{\text{ov}} = 
\begin{cases}
  D_\text{o} & \text{if $r_\text{o} < r < r_\text{o} + \alpha_{\text{ov}} H_\text{p}$} \\
  0 & \text{if $r > r_\text{o} + \alpha_{\text{ov}} H_\text{p}$}
\end{cases}
\end{array}
\label{step_os}
\end{equation}
where $D_\text{o}$ is determined just below the surface as explained above and $H_\text{p}$ is the pressure scale height at the boundary $r_\text{b}$. The parameter $\alpha_{\text{ov}}$ determines the extent of the overshooting region respectively. 
 \item \textbf{Exponential overshooting}: Motivated by 3-dimensional hydrodynamic simulations that suggest exponential decay of velocity field above the core, this prescription assumes an exponentially decreasing diffusion coefficient, thus reducing the efficiency of the chemical mixing away from the convective core. $D_{\text{ov}}$ in this case is given as
\begin{equation}
\begin{array}{c@{\qquad}c}
D_{\text{ov}} = D_\text{o} e^{-2(r-r_\text{o})/f_{\text{ov}}H_\text{p}}
\end{array}
\label{exp_os}
\end{equation}
where $r$ extends till $D_{\text{ov}}$ drops to a constant value of $D_{\text{ext}}$ in the radiative regions. The parameter $f_{\text{ov}}$ determines the inverse of the decay slope and thus the extent of the region, with larger $f_{\text{ov}}$ values resulting in larger overshooting region. 

 \item \textbf{Double exponential overshooting}: Apart from the two prescriptions discussed above, there is a third option available. It involves replacing the constant $( =D_{\text{ext}})$ diffusion profile in the radiative regions with a secondary exponential profile above the primary one. The variation in $D_{\text{ov}}$ is then as follows:
\begin{equation}
\begin{array}{c@{\qquad}c}
D_{\text{ov}} = 
\begin{cases}
  D_\text{o} e^{-2(r-r_\text{o})/f_\text{o} H_\text{p}} & \text{if $r_\text{o} < r < r_1$} \\
  D_1 e^{-2(r-r_1)/f_1H_\text{p}} & \text{if $r > r_1$}
\end{cases}
\end{array}
\label{double_exp_os}
\end{equation}
where $r_1$, the location of the switch, is determined by the value of the input parameter $D_1$. The decay of the second exponential profile is determined by $f_1$. For $f_\text{o} = f_1$, we get back the case of single exponential overshooting. The three different prescriptions are summarized in Fig. \ref{fig:os_prescription}.
\end{enumerate}

\begin{figure}
    \includegraphics[width = \columnwidth]{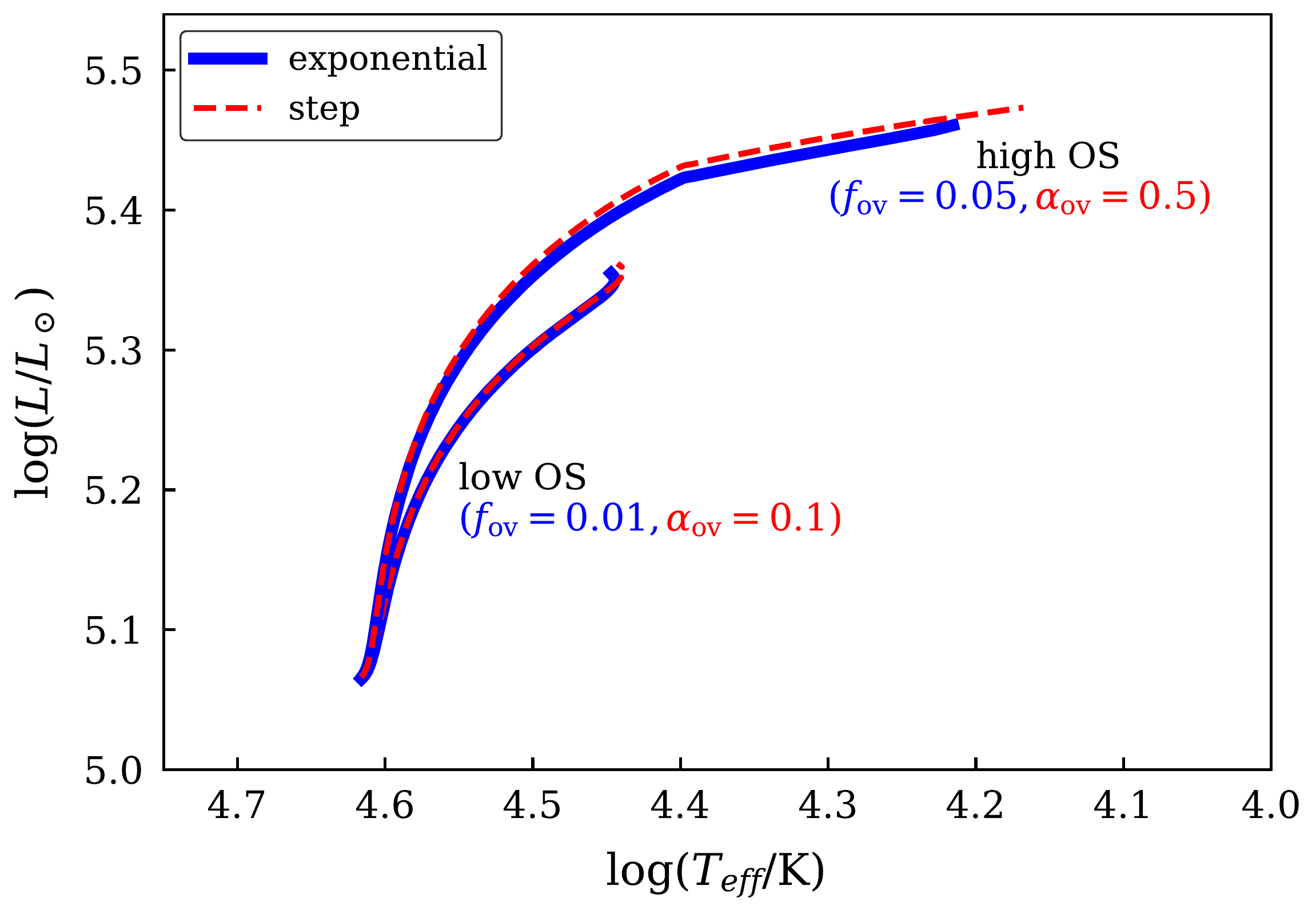}
    \caption{The main sequence evolution of a 30 $M_\odot$ star for two overshooting prescriptions - step (red dashed) and exponential (blue solid) - and different amounts of overshooting mixing above the hydrogen core - low and high. } 
    \label{fig:step_exp}
\end{figure}

Finally, we compare the step and exponential prescriptions and investigate the relation between their parameters that result in very similar main sequence evolution. To test this, we run models with two different values of step overshooting parameter $\alpha_{\text{ov}} = 0.1$ and $0.5$. We then compare their HRD evolution with models that have exponential profile of $f_{\text{ov}} = 0.01$ and $0.05$. Their main sequence evolution is shown in Fig. \ref{fig:step_exp}. Comparing the HRDs, models with an exponential overshooting has similar main sequence evolution as models with step overshooting for the approximate relation $f_{\text{ov}} \approx \alpha_{\text{ov}}/10$.

\section{MLT++ parameters}
\label{appendix:A}

\begin{figure}
\includegraphics[width=\columnwidth]{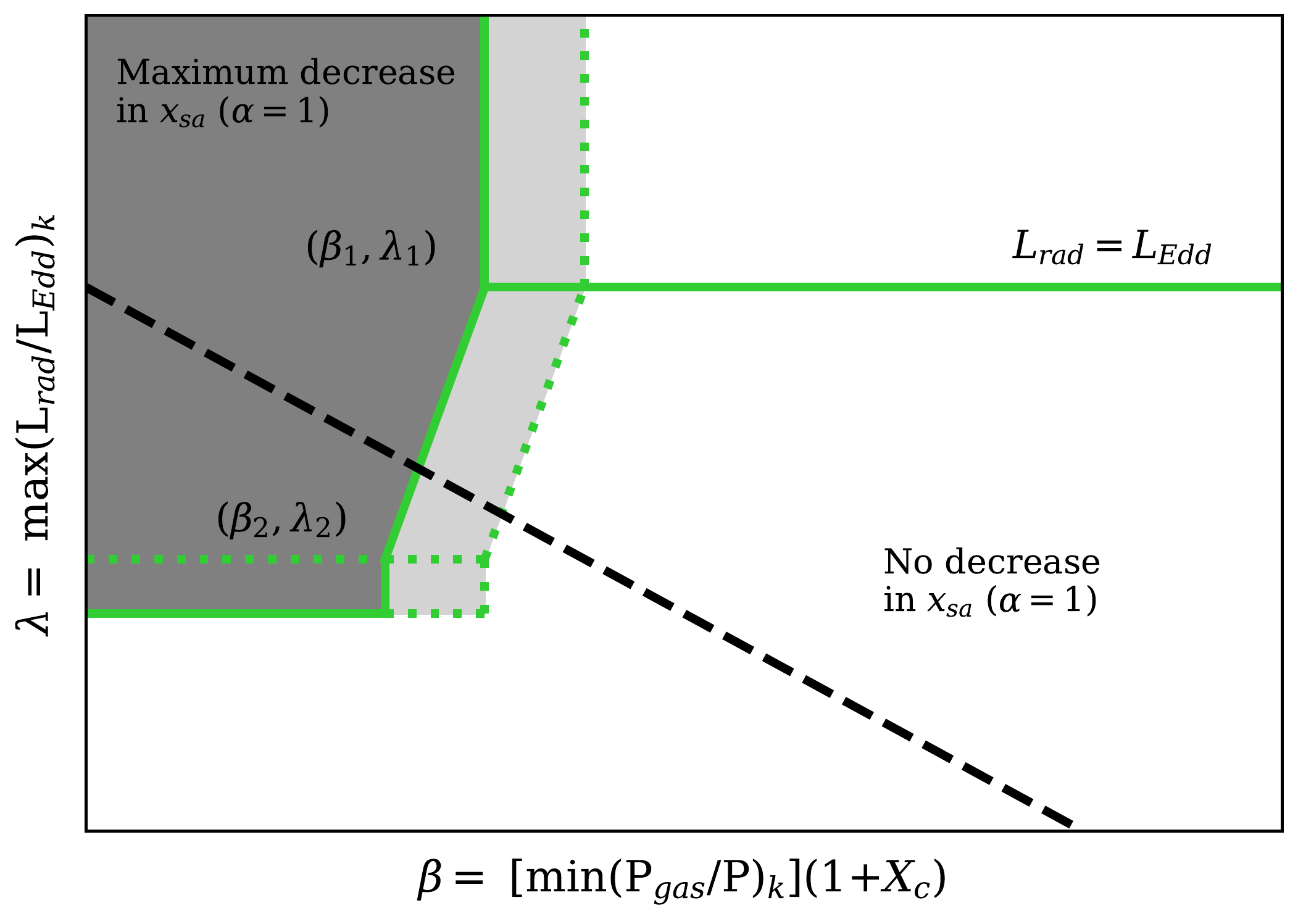}
    \caption{Typical parameter space of applicability of MLT++ as a function of minimum of $\beta = P_{\text{gas}}/P$ and maximum of $\lambda = L_{\text{rad}}/L_{\text{Edd}}$ inside the star where $k$ represents the cell index running from from the surface to the center. The shaded region receives full boost in efficiency, while the white region means no boost and no decrease in superadiabaticity. The transition between these two regions with $0<\alpha<1$ is of thickness 0.1 in both variables. The black dashed line is the condition of $\text{max}(\lambda) + \text{min}(\beta) = 1$ as discussed in Appendix \ref{appendix:B}. }
    \label{fig:mlt++}
\end{figure}

Here we first outline the MLT++ parameters that are used in our models, followed by the implementation of MLT++ routine in MESA in detail. MLT++ reduces the actual superadiabaticity $x_{\text{sa}}$ when it increases beyond the user defined threshold superadiabaticity  $x_{\text{sa,thresh}} = 10^{-4}$. The various parameters in MESA allow the user to choose the regions inside the star that is subjected to the treatment of MLT++. Fig. \ref{fig:mlt++} is used to determine when MLT++ is activated inside the star. Stars dominated by radiation pressure (low $\beta$) and close to their Eddington limit (high $\lambda$) that end up in the shaded region are subjected to maximum decrease in $x_{\text{sa}} (\alpha = 1)$. Models that lie in the white region, with high $\beta$ and low $\lambda$, are unaffected by the MLT++ routine and undergo no decrease in $x_{\text{sa}} (\alpha = 0)$. We use ($\beta_1$, $\lambda_1$) = (0.4, 1) and ($\beta_2$, $\lambda_2$) = (0.3, 0.5) for our models with a transition region of thickness ($\Delta\beta_1$, $\Delta\lambda_1$) = (0.1, 0.1) between the two regions. Switching on MLT++ in specific regions of the star affects its radial expansion, resulting in a hotter, more compact star that spends a small fraction of its helium burning time in the red supergiant regime. This causes the cutoff luminosity to shift to lower values.

As briefly discussed above, the decrease in superadiabaticity is decided by the location of the star in the $\beta-\lambda$ plane. Two conditions need to be satisfied for switching on MLT++ in inefficient convective zones of the star: actual superadiabaticity in these layers to exceed $x_\text{{sa, thresh}}$ (given by \texttt{gradT\_excess\_f1} in MESA) and the model to lie in the shaded region in Fig. \ref{fig:mlt++}. What remains to be addressed is the amount by which the actual superadiabaticity is affected when the aforementioned two conditions are satisfied. The actual temperature gradient is calculated as follows:
\begin{equation}
\begin{array}{c@{\qquad}c}
\nabla_{T, \text{new}} = f \nabla_T + (1-f) \nabla_{\text{ad}}
\end{array}
\label{gradT_MLT++}
\end{equation}
where $f$ sets the fraction by which the superadiabaticity is reduced. Setting a lower value of $f$ results in a larger decrease in the temperature gradient, pushing it closer to the adiabatic gradient, thus decreasing the superadiabaticity. For the case of full MLT++ boost ($\alpha = 1$), this fraction is set as $10^{-2}$ in our models (given by \texttt{gradT\_excess\_f2} in MESA). For a partial boost ($\alpha<1$), the value of $f$ is calculated as follows:
\begin{equation}
\begin{array}{c@{\qquad}c}
f = f_2 + (1-f_2)(1-\alpha) 
\end{array}
\label{f2_MLT++}
\end{equation}
Thus $f_2$ as well as $\alpha$ (although more indirectly) decide the amount by which MLT++ affects our models. The value of $f_2$ has minimal effect on our results provided a sufficiently low value is chosen. MESA also provides the option to limit the amount of decrease in superadiabaticity in one timestep. It also allows for smoothing of certain parameters such as $\alpha$ from one timestep to the next to prevent any sudden increases as the model evolves. 
\begin{equation}
\begin{array}{c@{\qquad}c}
\alpha_{t+dt} = (1 - f_{\text{age}})\alpha + f_{\text{age}} \alpha_t
\end{array}
\label{f2_MLT++}
\end{equation}
where $f_{\text{age}}$ is the time smoothing fraction that allows for a gradual increase (or decrease) in $\alpha$ as the model evolves, $\alpha_{t+dt}$ and $\alpha_{t}$ are the values of the efficiency boost corresponding to the new and old timesteps and $\alpha$ is the value obtained from the $\beta-\lambda$ plane. For the first timestep both $\alpha$ and $\alpha_t$ are obtained from the $\beta-\lambda$ plane. Thus, values of $\alpha < 1$ can be realised even in the shaded region, only truly reaching maximum decrease in superadiabaticities during evolution of stars with initial masses $\gtrsim 50 M_\odot$. For the initial masses considered in this study, $\alpha$ stays below 1 for most of the evolution \citep[Fig. 41]{MESA13}. There is also the option of completely switching off MLT++ during certain evolutionary stages such as the core hydrogen burning or beyond a certain minimum central helium mass fraction. These different options can affect individual models, but as long as mixing by MLT++ is turned on during the post-main sequence evolution and allows for appreciable changes in the superadiabaticities across timesteps, the overall results remains unchanged. In our models we do not employ any timestep restrictions on the changes in superadiabaticities, set the age fraction to 0.99 (a more smoother increase in $\alpha$ compared to the default value of 0.9) and switch on MLT++ throughout the evolution.


\begin{figure}
    \includegraphics[width = \columnwidth]{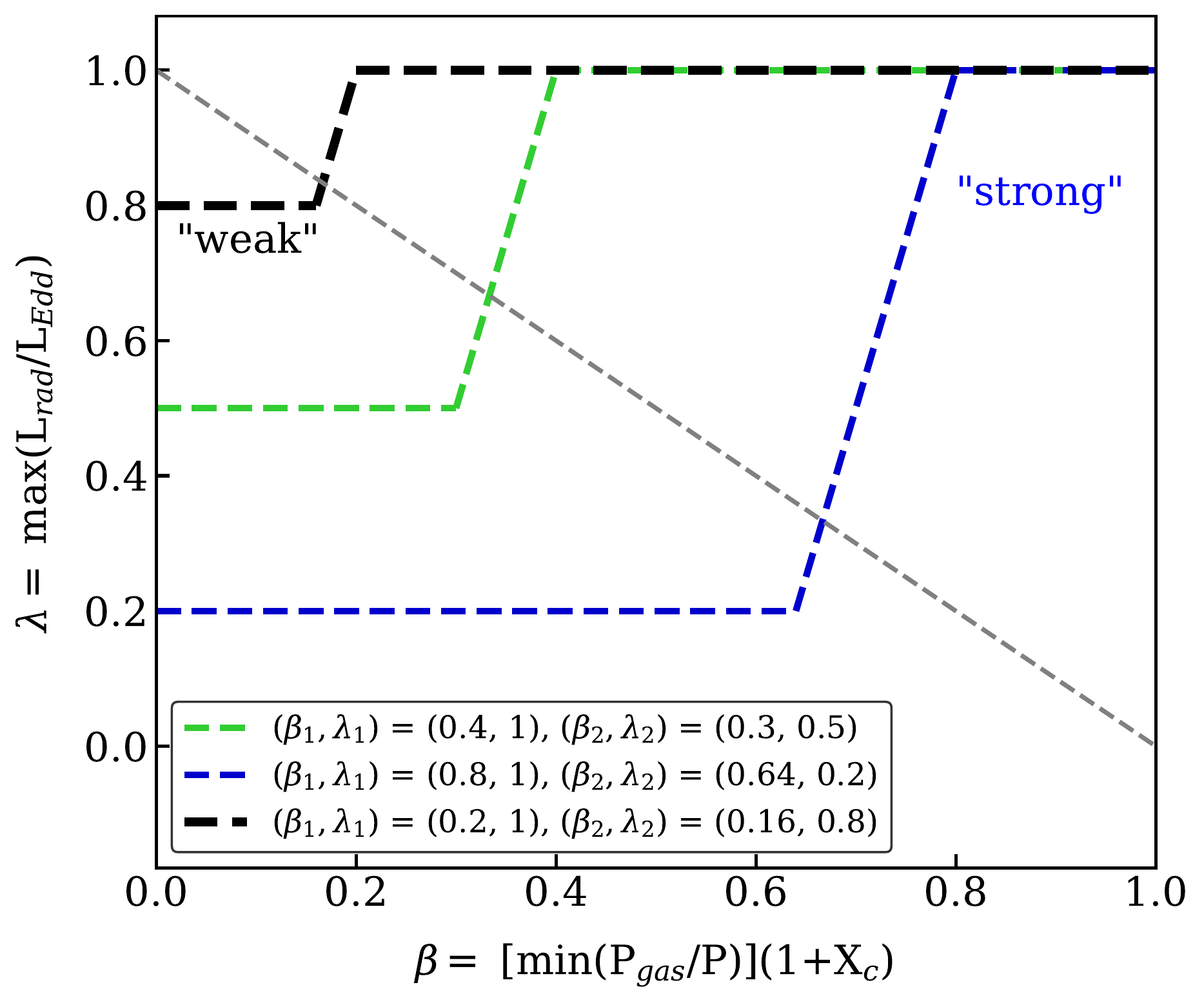}
    \caption{Parameters used for investigating the effects of different set of MLT++ parameters on the RSG luminosity limit. The grey dashed line is the condition of $\text{max}(\lambda) + \text{min}(\beta) = 1$ as discussed in Appendix \ref{appendix:B}.}
    \label{fig:MLT++PS1}
\end{figure}

In Sect. \ref{section:MLT++EFFECT}, we discussed the effects of 'weaker' and 'stronger' versions of MLT++ on our results. Here we define their parameters. Fig. \ref{fig:MLT++PS1} shows the different set of MLT++ parameters that we have used to investigate their effects on our results (Sect. \ref{section:MLT++EFFECT}). The green dashed line is the "standard" set of parameters used to reproduce our results. Depending on the definition of RSGs and their limit (we have chosen $\log (T_{\text{eff}}/\mathrm{K}) < 3.68$ and less than $5\%$ helium burning time spent above the limit), the parameter values used to reproduce the results can vary by $\approx \pm 0.1$. The "weaker" set of MLT++ parameter  that over-predicts the number of RSGs above the observed limit is shown in black, while the "stronger" set of parameter that hardly forms any RSGs is shown in blue.

\section{The extrema of $\beta$ and $\lambda$}
\label{appendix:B}

Here we briefly discuss the theory to understand the evolution of the parameters $\lambda$ and $\beta$ and the location of their extremes inside the star. Consider a massive star during its main sequence, which can be approximated by a $n=3$ polytrope despite having a convective core. The radiative pressure gradient is related to the radiative luminosity as follows
\begin{equation}
\begin{array}{c@{\qquad}c}
\dfrac{dP_{\text{rad}}}{dr} = \dfrac{4}{3}aT^3\dfrac{dT}{dr} = -\dfrac{\chi \rho}{4\pi c}\dfrac{L_{\text{rad}}(r)}{r^2}
\end{array}
\label{eq:radiative_pressure_grad}
\end{equation}
The condition of hydrostatic equilibrium inside the  star relates the  total pressure gradient required to support the inward gravitational force:
\begin{equation}
\begin{array}{c@{\qquad}c}
\dfrac{dP}{dr} = -\dfrac{Gm\rho}{r^2}
\end{array}
\label{eq:total_pressure_grad}
\end{equation}
Dividing Eq. \ref{eq:radiative_pressure_grad}  by Eq. \ref{eq:total_pressure_grad}  gives the following relation
\begin{equation}
\begin{array}{c@{\qquad}c}
\dfrac{dP_{\text{rad}}}{dP} = \dfrac{\chi L_{\text{rad}}(r)}{4\pi Gcm} = \dfrac{L_{\text{rad}}(r)}{L_{\text{Edd}}(r)} = \lambda(r)
\end{array}
\label{eq:dprad_dp}
\end{equation}
where $L_{\text{Edd}}(r)$ is the Eddington luminosity at radius $r$. Inside the star, well below the surface where one can assume $\beta \approx$ constant, we have $d\text{logP} \approx d\text{logP}_{\text{rad}}$. This results in a relation between the Eddington factor $\lambda$ and the ratio of gas pressure to total pressure $\beta$ at a given radius $r$:
\begin{equation}
\begin{array}{c@{\qquad}c}
\dfrac{dP_{\text{rad}}}{dP} \approx \dfrac{P_{\text{rad}}}{P} \implies \dfrac{L_{\text{rad}}(r)}{L_{\text{Edd}}(r)} + \dfrac{P_{\text{gas}}}{P} \approx 1
\end{array}
\label{eq:relate_beta_lambda}
\end{equation}

\begin{figure}
    \includegraphics[width = \columnwidth]{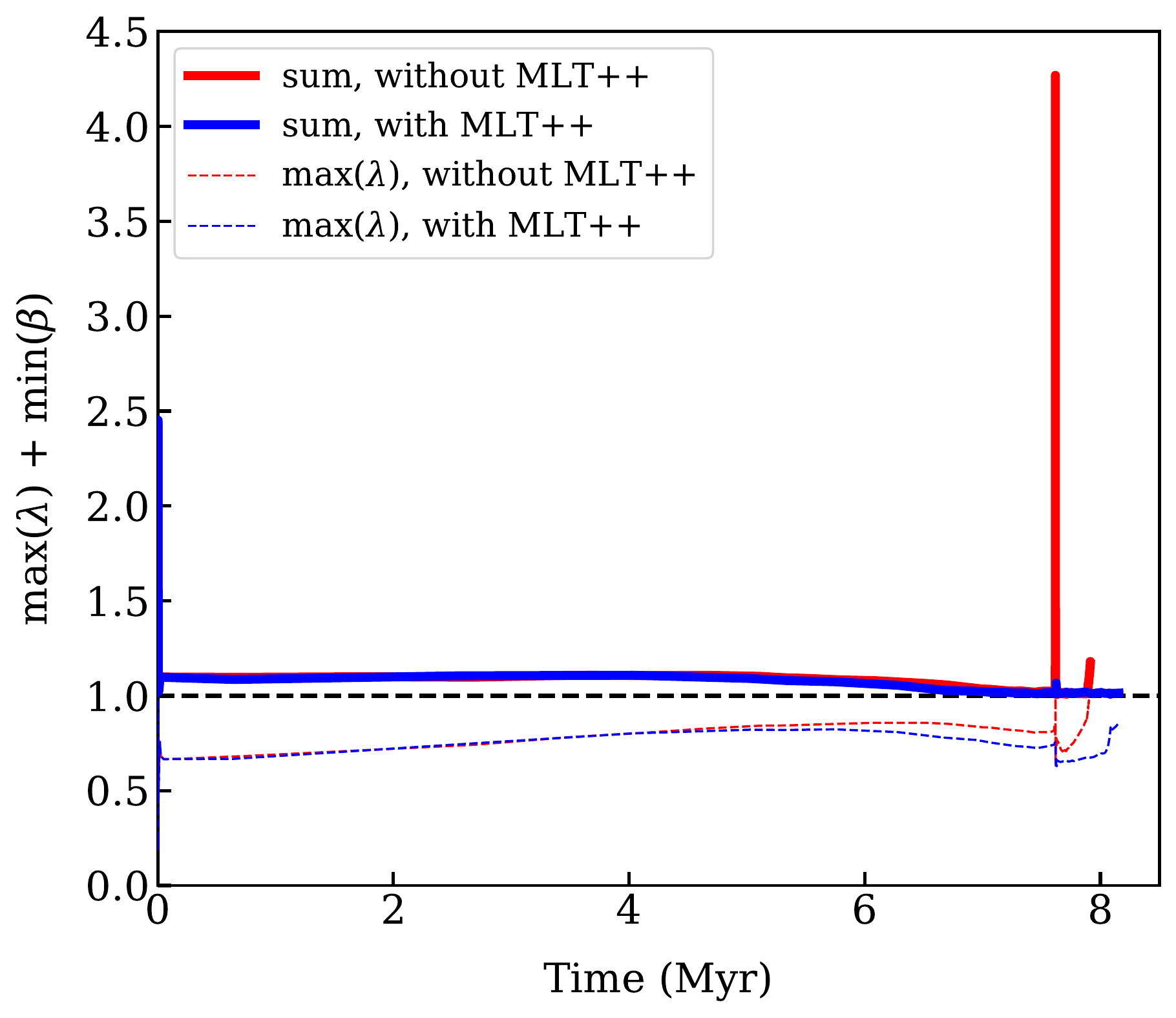}
    \caption{Variation of sum of max($\lambda$) and min($\beta$) for a 25 $M_\odot$ star, with (blue) and without MLT++ (red), as  it evolves from ZAMS to the end of helium burning. Super-Eddington conditions (red dashed line) are achieved in the non-MLT++ models at the beginning (near hydrogen bump) of core helium burning and at about half way through helium burning (near HeII bump). The black dashed line represents the unity condition from Eq. \ref{eq:relate_beta_lambda}.} 
    \label{fig:beta+lambda}
\end{figure}

Again, this is only true inside the star where one can approximately assume that changes in $\beta$ are gradual around their extremum, and one can assume $\beta$ as a constant. Any rapid changes in $\beta$ effectively violating this constancy result in the sum of the two quantities on the left in Eq. \ref{eq:relate_beta_lambda} deviating away from unity. If the maximum value of $\lambda$ and the minimum $\beta$ inside the star occurs at roughly the same radius $r$, then we can write max($\lambda$)+min($\beta$) = 1. This only holds true if the two extrema occur at the same location inside the star.

\begin{figure}
    \includegraphics[width = \columnwidth]{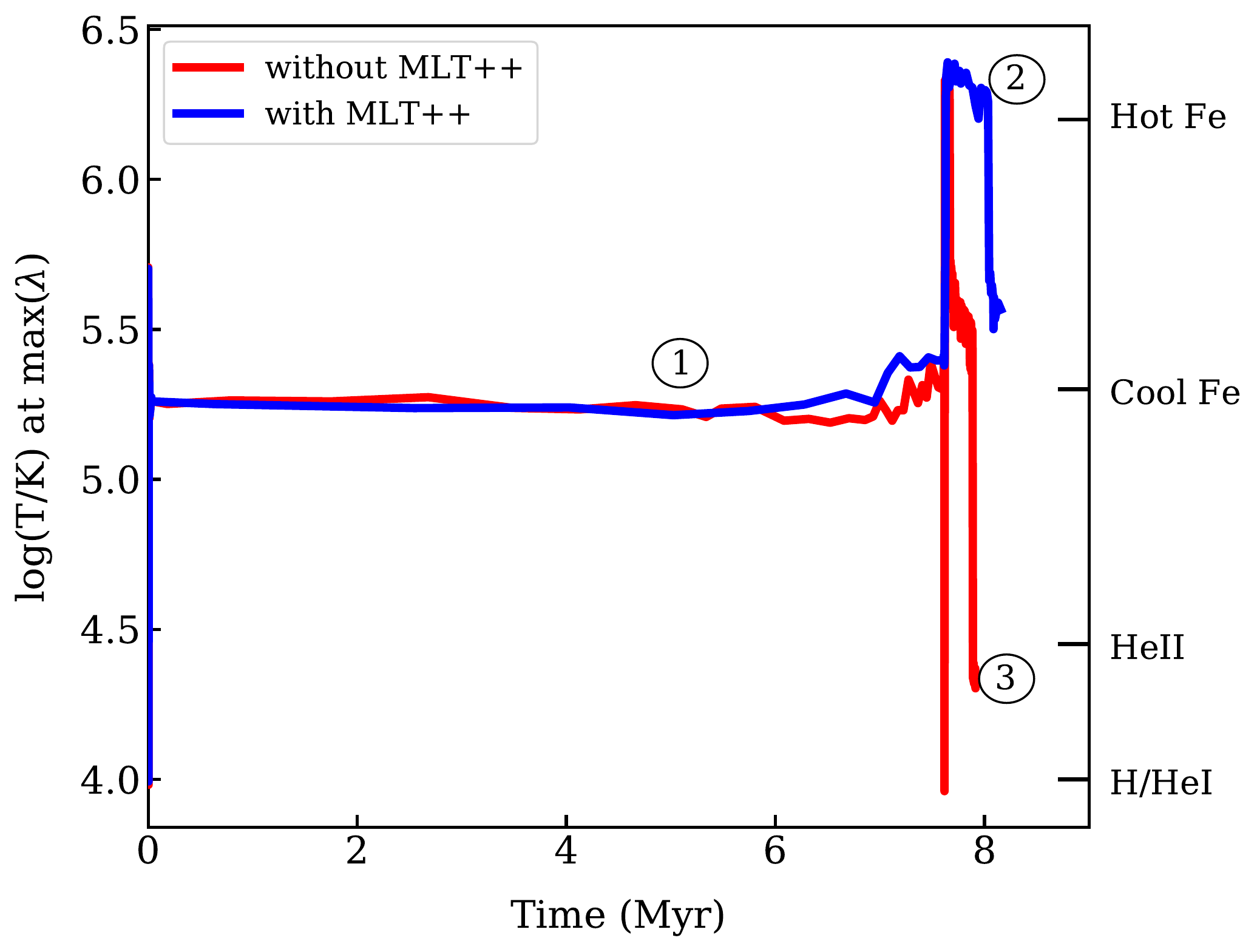}
    \caption{Variation of the temperature location of the maximum $\lambda$ inside the star for models with (blue) and without MLT++ (red). The locations of the different opacity bumps are also marked on the right. The points marked 1, 2 and 3 are explained in the text.} 
    \label{fig:Temp_loc}
\end{figure}

The condition in Eq. \ref{eq:relate_beta_lambda}, if strictly satisfied around the location of maximum $\lambda$ inside the star, must coincide with the location of minimum $\beta$. Fig. \ref{fig:beta+lambda} shows the evolution of the sum of maximum $\lambda$ and minimum $\beta$ inside a star with initial mass of 25 $M_\odot$, with (blue solid) and without MLT++ (red solid), from the start of core hydrogen burning to the end of core helium burning (the case without MLT++ is evolved only till $Y_c \approx 0.33$ due to timesteps problems). Note that all other inputs are fixed except for the usage of MLT++. The dashed lines show the contribution of max($\lambda$) to the sum with (blue dashed) and without MLT++ (red dashed). In Fig. \ref{fig:beta+lambda}, we see that the sum remains close to unity during the main sequence, only departing from the aforementioned condition during helium burning when cooler opacity bumps are available where $\lambda(r)$ itself can easily cross unity. The red dashed line (max($\lambda$) without MLT++) crosses unity during core helium burning where the maximum $\lambda$ occurs near the outer opacity bumps of hydrogen and helium. As seen in Fig. \ref{fig:beta+lambda}, the MLT++ routine effectively suppresses the contribution of radiative luminosity in the the super-Eddington layers by making convection more efficient. While the model without MLT++ has a maximum $\lambda$ value as large as $\approx 4$ (occurs near the hydrogen opacity bump), the model with MLT++ keeps the layers sub-Eddington throughout the evolution of the star (blue dashed line). 

In Fig. \ref{fig:Temp_loc}, we show the temperature at which the maximum $\lambda$ occurs inside the star, along with the temperatures of the prominent opacity bumps. For the model without MLT++, the maximum $\lambda$ remains close to the cool Fe bump throughout the main sequence (1), and shifts to the HeII bump (3) when $Y_c\approx 0.4$ and remains there till the end of the model run. The instances when the maximum occurs close to the hydrogen and helium bump coincides exactly with the instances when these layers go super-Eddington close to the surface (compare with Fig. \ref{fig:beta+lambda}). The model with MLT++ on the other hand, that suppresses the Eddington parameter, has its location of maximum $\lambda$ shift inwards staying close to the hot iron bump (2). Thus, MLT++ acts on the superadiabatic regions that develop super-Eddington conditions, increases the efficiency of convection, effectively suppresses the $\lambda$ in these regions and pushes the maximum $\lambda$ inside the star towards the hotter opacity bumps.

\bsp	
\label{lastpage}
\end{document}